\documentclass[twocolumn,epjc3]{svjour3}         

\RequirePackage[T1]{fontenc}
\smartqed  
\usepackage[utf8]{inputenc}
\RequirePackage{graphicx}
\RequirePackage{mathptmx}      
\RequirePackage{flushend}
\RequirePackage[numbers,sort&compress]{natbib}
\usepackage{graphics}
\usepackage{xurl}
\usepackage{float}
\usepackage[switch,columnwise]{lineno}
\usepackage{testhyphens}
\usepackage{hyphenat}
\usepackage{mathptmx}
\usepackage{color}
\usepackage{graphicx} 

\input{my_ds_macro.def}
\input{se.def}

\newcommand{\APC}{APC, Universit\'e Paris Diderot, CNRS/IN2P3, CEA/Irfu, Obs de Paris, USPC, Paris 75205, France}
\newcommand{\AQLNGS}{INFN Laboratori Nazionali del Gran Sasso, Assergi (AQ) 67100, Italy}
\newcommand{\AQGSSI}{Gran Sasso Science Institute, L'Aquila 67100, Italy}

\newcommand{\Augustana}{Physics Department, Augustana University, Sioux Falls, SD 57197, USA}
\newcommand{\Belgorod}{Radiation Physics Laboratory, Belgorod National Research University, Belgorod 308007, Russia}

\newcommand{\CAUniPHY}{Physics Department, Universit\`a degli Studi di Cagliari, Cagliari 09042, Italy}
\newcommand{\CAINFN}{INFN Cagliari, Cagliari 09042, Italy}

\newcommand{\CTLNS}{INFN Laboratori Nazionali del Sud, Catania 95123, Italy}
\newcommand{\ENUniCEE}{Engineering and Architecture Faculty, Universit\`a di Enna Kore, Enna 94100, Italy}

\newcommand{\FNAL}{Fermi National Accelerator Laboratory, Batavia, IL 60510, USA}

\newcommand{\GEUni}{Physics Department, Universit\`a degli Studi di Genova, Genova 16146, Italy}
\newcommand{\GEINFN}{INFN Genova, Genova 16146, Italy}

\newcommand{\Hawaii}{Department of Physics and Astronomy, University of Hawai'i, Honolulu, HI 96822, USA}
\newcommand{\Houston}{Department of Physics, University of Houston, Houston, TX 77204, USA}
\newcommand{\IHEP}{Institute of High Energy Physics, Beijing 100049, China}

\newcommand{\JINR}{Joint Institute for Nuclear Research, Dubna 141980, Russia}
\newcommand{\Krakow}{M. Smoluchowski Institute of Physics, Jagiellonian University, 30-348 Krakow, Poland}
\newcommand{\Kurchatov}{National Research Centre Kurchatov Institute, Moscow 123182, Russia}

\newcommand{\LNFINFN}{INFN Laboratori Nazionali di Frascati, Frascati 00044, Italy}

\newcommand{\LPNHE}{LPNHE, CNRS/IN2P3, Sorbonne Universit\'e, Universit\'e Paris Diderot, Paris 75252, France}
\newcommand{\Manchester}{The University of Manchester, Manchester M13 9PL, United Kingdom}
\newcommand{\MEPhI}{National Research Nuclear University MEPhI, Moscow 115409, Russia}

\newcommand{\MIINFN}{INFN Milano, Milano 20133, Italy}

\newcommand{\MIUni}{Physics Department, Universit\`a degli Studi di Milano, Milano 20133, Italy}
\newcommand{\MSU}{Skobeltsyn Institute of Nuclear Physics, Lomonosov Moscow State University, Moscow 119234, Russia}
\newcommand{\NAINFN}{INFN Napoli, Napoli 80126, Italy}
\newcommand{\NAUniPHY}{Physics Department, Universit\`a degli Studi ``Federico II'' di Napoli, Napoli 80126, Italy}

\newcommand{\Petersburg}{Saint Petersburg Nuclear Physics Institute, Gatchina 188350, Russia}
\newcommand{\PGUniCBB}{Chemistry, Biology and Biotechnology Department, Universit\`a degli Studi di Perugia, Perugia 06123, Italy}
\newcommand{\PGINFN}{INFN Perugia, Perugia 06123, Italy}
\newcommand{\PIINFN}{INFN Pisa, Pisa 56127, Italy}
\newcommand{\PIUniPHY}{Physics Department, Universit\`a degli Studi di Pisa, Pisa 56127, Italy}
\newcommand{\PNNL}{Pacific Northwest National Laboratory, Richland, WA 99352, USA}
\newcommand{\Princeton}{Physics Department, Princeton University, Princeton, NJ 08544, USA}

\newcommand{\RMTreINFN}{INFN Roma Tre, Roma 00146, Italy}
\newcommand{\RMTreUni}{Mathematics and Physics Department, Universit\`a degli Studi Roma Tre, Roma 00146, Italy}
\newcommand{\RMUnoINFN}{INFN Sezione di Roma, Roma 00185, Italy}
\newcommand{\RMUnoUni}{Physics Department, Sapienza Universit\`a di Roma, Roma 00185, Italy}

\newcommand{\Stanford}{Physics Department, Stanford University, Stanford, CA 94305, USA}

\newcommand{\UCDavis}{Department of Physics, University of California, Davis, CA 95616, USA}
\newcommand{\UCLA}{Physics and Astronomy Department, University of California, Los Angeles, CA 90095, USA}
\newcommand{\UMass}{Amherst Center for Fundamental Interactions and Physics Department, University of Massachusetts, Amherst, MA 01003, USA}

\newcommand{\USP}{Instituto de F\'isica, Universidade de S\~ao Paulo, S\~ao Paulo 05508-090, Brazil}
\newcommand{\VTech}{Virginia Tech, Blacksburg, VA 24061, USA}
\newcommand{\london}{Physics, Kings College London, Strand, London WC2R 2LS, United Kingdom}

\newcommand{\CPPM}{Centre de Physique des Particules de Marseille, Aix Marseille Univ, CNRS/IN2P3, CPPM, Marseille, France}
\newcommand{\SDakota}{School of Natural Sciences, Black Hills State University, Spearfish, South Dakota 57799, USA}
\newcommand{\AstroCeNT}{AstroCeNT, Nicolaus Copernicus Astronomical Center of the Polish Academy of Sciences, 00-614 Warsaw, Poland}
\newcommand{\UCRiverside}{Department of Physics and Astronomy, University of California, Riverside, CA 92507, USA}
\newcommand{\Oxford}{University of Oxford, Oxford OX1 2JD, United Kingdom}
\newcommand{\Washington}{Center for Experimental Nuclear Physics and Astrophysics, and Department of Physics, University of Washington, Seattle, WA 98195, USA}
\newcommand{\UCAS}{University of Chinese Academy of Sciences, Beijing 100049, China}
\newcommand{\Williams}{Williams College, Physics Department, Williamstown, Massachusetts 01267, USA}

\renewcommand{\thanksref}[1]{\nolinebreak\textsuperscript{\ref{#1}}\nolinebreak}

\graphicspath{{figs/}}

\journalname{Eur. Phys. J. C}

\makeatletter
\usepackage{url}
\usepackage{hyperref}

\begin{document}
\runningpagewiselinenumbers
\sloppy   

\title{Characterization of spurious-electron signals in the double-phase argon TPC of the \DSf\ experiment}
\author{The \DSf\ Collaboration$^\text{\normalfont a,1}$}

\thankstext{e1}{e-mail: eberzin@stanford.edu}
\institute{See back for author list \label{addr1}}

\date{\today}

\maketitle

\abstract{
Spurious-electron signals in dual-phase noble-liquid time projection chambers have been observed in both xenon and argon Time Projection Chambers (TPCs). This paper presents the first comprehensive study of spurious electrons in argon, using data collected by the \DSf\ experiment at the INFN Laboratori Nazionali del Gran Sasso (LNGS). Understanding these events is a key factor in improving the sensitivity of low-mass dark matter searches exploiting ionization signals in dual-phase noble liquid TPCs.

We find that a significant fraction of spurious-electron events, ranging from 30 to 70\% across the experiment's lifetime,  are caused by electrons captured from impurities and later released with delay time constants on the orders of \SI{5}{\ms} and \SI{50}{\ms}. The rate of spurious-electron events is found to correlate with the operational condition of the purification system and the total event rate in the detector. Finally, we present evidence that multi-electron spurious electron events may originate from photo-ionization of the steel grid used to define the electric fields. These observations indicate the possibility of reduction of the background in future experiments and hint at possible spurious electron production mechanisms.



}

\section{Introduction}
\label{sec:intro}
The direct search for  Weakly Interacting Massive Particle (\WIMP) Dark Matter (\DM) is one of the most active areas of research in astroparticle physics. 
Two key parameters that determine low-mass (\SI{<10}{GeV/c^2}) DM sensitivity with dual-phase noble-liquid time projection chambers (\TPCs)  are the minimum detectable recoil energy and the background rate. 

Heavy WIMP (\SI{>10}{GeV/c^2}) interactions in a dual-phase TPC are characterised by a prompt scintillation signal (\SOne),\ followed by a delayed signal (\STwo)\ produced by ionization electrons. These electrons drift upward in the liquid due to a uniform electric field and are extracted into a thin gaseous layer (gas pocket), where they induce electroluminescence.
As discussed in Refs.~\cite{Agnes:2017ck,Agnes:2018hvf}, \SOne\ and \STwo\ have different pulse shapes. 
In liquid argon (\LAr), \SOne\ rises in a few nanoseconds and falls as a double exponential, with mean lifetimes \mbox{$\tau_1\sim\SI{6}{\ns}$} and \mbox{$\tau_2=\SI{1.4\pm0.1}{\micro\second}$}~\cite{adhikariLiquidargonScintillationPulseshape2020}, 
and a prompt fraction (light collected in the first \SI{90}{\ns} over total) of approximately \SI{30}{\percent} (\SI{70}{\percent})
for electronic (nuclear) recoils.  
This difference in S1 pulse shape enables efficient pulse shape discrimination (PSD) between electronic and nuclear recoils~\cite{adhikariPulseshapeDiscriminationLowenergy2021}.
In the \DSf\ configuration (described in \refsec{detector}), \STwo\ has a characteristic rise-time of \SI{1.2\pm0.5}{\micro\second} and falls exponentially with a \SI{3.43\pm0.06}{\micro\second} time constant~\cite{Agnes:2018hvf}, and lasts for a few tens of \si{\mu s}. 
Detecting both \SOne\ and \STwo\ allows for the 3D position reconstruction of interaction vertices, enabling background rejection through fiducialization and multiple-scatter rejection.

Low-mass \WIMP\ searches, focused on \SI{<10}{\GeV\per\square c} masses, have been performed using primarily the ionization signal to detect interesting events (\stwoonly\ analysis).
While the \SOne\ photon detection efficiency is limited by the detector's light collection efficiency, extracted electrons generate an \STwo\ signal that can be detected with high efficiency due to the secondary emission of photons in the gas pocket, caused by the acceleration of electrons in the high field region (electroluminescence).
In \DSf\ this amplification was $\SI{23(1)}{\pe\ \text{(photoelectrons)}\ \per\el}$~\cite{Agnes:2018fg}. 
As a result, lower energies can be accessed in the \stwoonly\ channel than when both \SOne\ and \STwo\ are required. 
The  \stwoonly\  approach has been used in several light \DM\ searches~\cite{Agnes:2018fg,Agnes:2018ft,Aprile_2014,PhysRevLett.123.251801,PhysRevLett.107.051301,PhysRevLett.109.021301,xenoncollaborationSearchLightDark2025}, and it is proposed for future searches for light \DM~\cite{agnesSensitivityProjectionsDualphase2022, noauthor_darkside-20k_2024} and Coherent Elastic Neutrino-Nucleus Scattering (CE$\nu$NS) bursts from supernova neutrinos~\cite{Agnes_2021a}.
\stwoonly\ analyses have also been used to search for CE$\nu$NS from reactor neutrinos~\cite{red-100_collaboration_first_2025} and solar neutrinos~\cite{pandaxcollaborationFirstIndicationSolar2024}.

All  \stwoonly\ analyses,   with both   argon~\cite{Agnes:2018fg,Agnes:2018ft,DarkSide-50:2022hin,DarkSide-50:2022qzh,DarkSide:2022dhx} and xenon~\cite{Akimov_2016,Aprile_2014,EDWARDS200854,zeplin11,Sorensen:2017ymt,Sorensen:2017kpl,PhysRevLett.107.051301,PhysRevLett.123.251801,bodniaElectricFieldDependence2021,kopecCorrelatedSingleFewelectron2021a,sorensenQuartzFluorescenceBackgrounds2025}, have seen excess rates of signals corresponding to a few extracted electrons, herein referred to as Spurious Electrons (SEs), inconsistent with radioactive background models (illustrated and discussed below, in Fig.~\ref{fig:se_nedist}). 
Their origin is not completely understood, and their presence limits the sensitivity of these experiments.
For instance, for the \DSf\ \stwoonly\ light DM searches~\cite{DarkSide-50:2022hin,DarkSide-50:2022qzh,DarkSide:2022dhx}, only events at energies where the SE background was found to be negligible were retained, thus restricting sensitivity to WIMP masses above \SI{1.2}{\GeV\per\square\c}.
Understanding the nature of these events is therefore the key to improving the sensitivity of low-mass  WIMP searches.

This paper presents the first systematic study of SE events in LAr, using data collected by the \DSf\ detector located at LNGS. The findings from this study are compared to similar findings with xenon detectors.

In \refsec{detector}, we describe the \DSf\ detector. 
Section~\ref{sec:analysis} describes the data selection and classification of event types, including SE selection cuts and event rate trends.
These trends are correlated with measures of the detector purity in \refsec{purity}.
Correlations of SEs with other events and with detector parameters are explored in \refsec{correlations}. 
The electron multiplicity in the SE events is studied in Section~\ref{sec:ele_multi}. 
Finally, \refsec{concl} places results in a broader context and discusses possible mechanisms of SEs production/emission, followed by a summary in \refsec{summary}.

\section{The \DSf\ detector}
\label{sec:detector}

\DSf\ \LArTPC\ has a cylindrical target defined by a \DSfLArBelowMeshHeight\ diameter and a \DSfLArBelowMeshHeight\ height (at \LAr\ temperature) and consisting of  \DSfActiveMass\ of underground argon (\UAr) with an \ce{^39Ar} concentration at least \DSfUArDepletion-times lower than in atmospheric argon~\cite{AcostaKane:2008im,Xu:2015do}. 
\SOne\ and \STwo\ photons are detected with two arrays of 19 \DSfPMTSize\ diameter  Hamamatsu R11065 photomultiplier tubes (\PMTs), which view the \UAr\ from above and below through fused-silica windows. The windows are coated on both faces with \SI{15}{\nano\meter}-thick transparent and conductive indium tin oxide (ITO). 
These coatings form the \TPC's grounded anode (top) and negative high-voltage cathode (bottom), while their outer faces are held at the mean \PMT\ photocathode potential. 

The cylindrical wall is a \SI{2.54}{\cm}-thick polytetrafluoroethylene (PTFE) reflector fabricated with a modified annealing cycle to increase its reflectivity. 
The \TPC's inner surfaces are coated with tetraphenyl butadiene (\TPB), a wavelength shifter that absorbs \LAr's \SI{128}{\nano\meter} scintillation photons and re-emits visible photons (peaked at \SI{420}{\nano\meter}), detected by the \PMTs\ with \SI{35}{\percent} peak quantum efficiency.
Ionization electrons drift towards the \LAr\ surface at  \espeedbelowmesh\ due to a \DSfDriftField\ drift field, with a maximum drift time of \DSfDriftTimeMax.
An etched \SI{50}{\micro\meter}-thick stainless steel grid positioned \DSfLArAboveMeshHeight\ (at \LAr\ temperature) below the liquid surface creates a \SIrange[range-units = single,range-phrase=--]{2.8}{3.7}{\kilo\volt\per\cm} extraction field in the liquid above the grid extracting electrons into the gaseous argon (\GAr) with \DSfElectronExtractionEfficiency\ efficiency~\cite{Bondar:2009gh} and a \SIrange[range-units = single,range-phrase = --]{4.2}{5.6}{kV/cm} electroluminescence field in the gas pocket produciung \STwo\ via electroluminescence. These variations of the field strengths are caused by local variations in the gas pocket thickness ({\it e.g.} due to sagging of anode window) or shape deformation of the grid~\cite{zhuStudyArgonElectroluminescence2018}.

To maintain and improve purity, argon is extracted from the cryostat in the gas phase by a circulation pump at a nominal rate of \SI{\sim28}{SLPM} (\SI{2.69}{kg/h}).
The \GAr\ passes through a hot getter (SAES Monotorr PS4-MT50-R-2~\cite{saes}), before passing through a cold charcoal radon trap, cooled by the \GAr. 
Immediately before entering the cold trap, the inlet \GAr\ is cooled by outlet nitrogen gas from a \LAr\ condenser through a heat-exchanger, decreasing the required cooling power. 
The temperature of the cold trap is correlated with the \GAr\ flow rate. 
Condensed \LAr\ is directly injected into the \TPC\ by gravity. See Ref.~\cite{Agnes_2024_longstability} for the schematic of the \DSf\ cryogenic system.

A hardware event trigger occurs when signals in \DSfTriggerPMTsNumberThreshold\ or more PMT channels exceed \SI{\sim0.6}{photoelectrons} (PE) within a \DSfTriggerWindow\ window. Subsequent triggers are inhibited for \SI{810}{\us}, and waveform data are recorded from all 38 \PMTs\ for \SI{440}{\us} starting \SI{10}{\us} before the trigger.
The data reported here were acquired between April 2015 and February 2018.

\section{Data selection and classification}
\label{sec:analysis}

\begin{figure*}[htb]
\begin{center}
\includegraphics[
trim=0.5cm 20cm 3cm 0.cm,
width=0.7\textwidth]{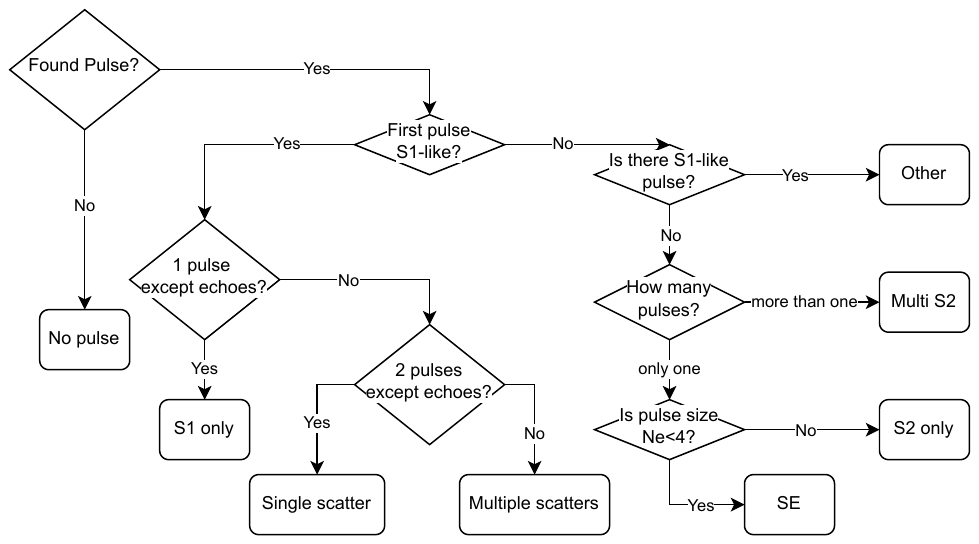}
 \caption{Event categorization scheme. After the basic requirements, all events are categorized in one of those categories.}
 \label{fig:evt_classification}
\end{center}
\end{figure*}

All triggered events are processed with the same low-level reconstruction software used in Ref.~\cite{Agnes:2015gu}. 
A pulse-finding algorithm is applied to the digitized data, including the pre-trigger window. Each identified pulse is characterized as \SOne\ or \STwo\ based on a calculated \FNine\ parameter, defined as the fraction of light detected in the pulse's first \WindowFNine: \SOne\ pulses have \FNine\num{>0.15}, whereas slower-rising \STwo\ pulses tend to have lower \FNine\ values. The efficiency of the pulse finding algorithm is essentially \DSfPulseFindingEfficiency\ for \STwo\ signals larger than \DSfPulseFindingEfficiencyThreshold~\cite{Agnes:2015gu}.  
No corrections accounting for the trigger and pulse-finding inefficiency are applied. 

Due to the observed radial variation in electroluminescence yield~\cite{Agnes:2018fg}, the trigger and pulse-finding efficiencies for single electron events depend on the events' radial positions. \STwo-like pulses are corrected for this variation using the radial position of the \PMT\ channel recording the most light, as in Ref.~\cite{Agnes:2018ft}. After normalizing for \PMT\ gain, this corrected \STwo\ size is referred to as the number of photoelectrons (\nPE).
For \STwo-like pulses, \nPE\ is converted to the number of extracted electrons (\Ne) by dividing by the single electron light yield under the central PMT:
$\Ne=\nPE/g_{2}$, where $g_{2}=\SI{23(1)}{\pe\per\el}$.

To characterize SEs, we define an event classification scheme. 
In the following section, all quoted uncertainties are purely statistical unless otherwise stated.

\subsection{Pulse classification}
Identified pulses are classified by \FNine\ and \nPE\ into two categories: \SOne-like and \STwo-like. 
``Echoes'', S2-like pulses produced by photo-ionization electrons released from the cathode when struck by \SOne\ or \STwo\ photons, are not counted in the total number of pulses. 
These echoes follow a preceding \SOne\ or \STwo\ pulse by exactly the maximum drift time for electrons in the \TPC\ and are characterized in Ref.~\cite{agnesStudyEventsPhotoelectric2022a}.


\begin{table*}[htb]
    \centering
    \caption{Summary of event classification categories}
    \begin{tabular}{l|l|l}\hline\hline
        Classification & Description & Examples \\\hline 
        No pulse & No pulses found & Noise trigger, low \Ne\ event near walls \\ 
        S1-only & \SOne\ identified; no \STwo\ found & Cherenkov, surface events \\ 
        Single-scatter & One \SOne; one \STwo\ pulse found & Particles scattering once in LAr, \SOne\ pileup \\ 
        Multi-scatter & One \SOne; multiple \STwo\ pulses identified & \grs\ scattering multiple times in LAr, random pileup \\ 
        Multi-\STwo & No \SOne; multiple \STwo\ pulses & Multi-scatters with low enough drift time (\tdrift) or sub-threshold \SOne \\
        \STwo\ only & No \SOne; one \STwo\ found with $\Ne\geq4$ & Single-scatter events with \SOne\ below threshold \\ 
        SE & No \SOne; one \STwo\ found with $\Ne<4$ & Delayed electrons \\ 
        Other & \STwo\ pulse first, followed by at least one \SOne\ pulse & \STwo-triggered events with random pile-up \\ 
        \hline\hline 
    \end{tabular}
    \label{tab:pulse_classes}
\end{table*}

\subsection{Event classification}

All events are required to satisfy the following three basic requirements as in Ref.~\cite{Agnes:2018ep}:
(1) data are present for all \TPC\ channels; (2) baselines for the digitized waveforms are successfully found in all \TPC\ channels; and (3) the event occurs at least \DSfDdMinLivetimeCut\ after the end of the inhibit window of the previous trigger (that is, at least \SI{1.21}{\milli\second} after the previous trigger).  
The last requirement removes events whose \SOne\ might have occurred during the inhibit window, and are subsequently triggered by the following \STwo\  pulse.

All events passing these cuts are classified into one of the categories described in Table~\ref{tab:pulse_classes}. 
If the pulse-finder finds no pulses, the event is categorized as ``No pulse'', as happens when events are triggered by a small signal that the pulse-finder fails to identify. 
Events in which at least one pulse is found are separated by whether the first pulse is \SOne-like or \STwo-like. 
Events with an \SOne-like first pulse are further divided into three categories depending on the number of subsequent S2 pulses: ``\SOne-only'' for events with only one pulse, ``Single-scatter'' for events with two pulses, and ``Multiple scatters'' for events with more than two pulses.  
Events with only \STwo-like pulses are divided into three categories based on the number of pulses: ``Multi-S2'' for events with more than one pulse, ``\STwo-only" for events with one pulse with $\Ne \geq 4$ (based on the \Ne\ distribution in Fig.~\ref{fig:se_nedist}), and ``SEs'' for events with one \STwo-like pulse with $\Ne < 4$. 
Events with an \STwo-like first pulse and at least one \SOne-like pulse are categorized as ``Other''. 
Figure~\ref{fig:evt_classification} shows a flow chart for the event classification.

%
\begin{figure*}
\begin{center}
\includegraphics[width=0.75\textwidth]{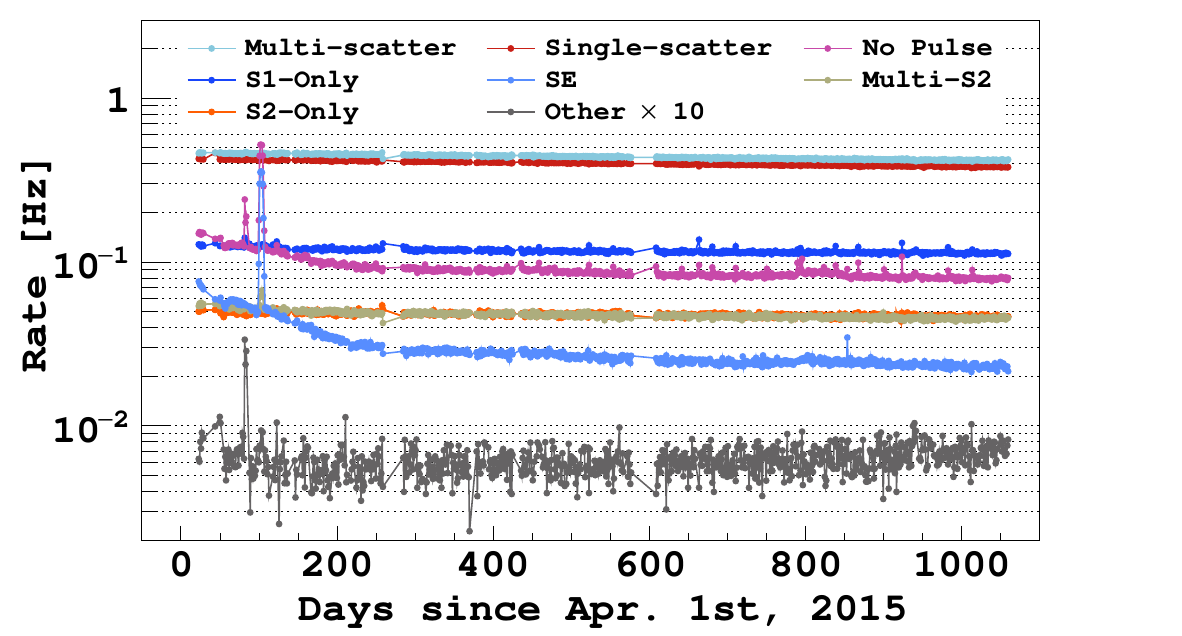}
\includegraphics[width=0.75\textwidth]{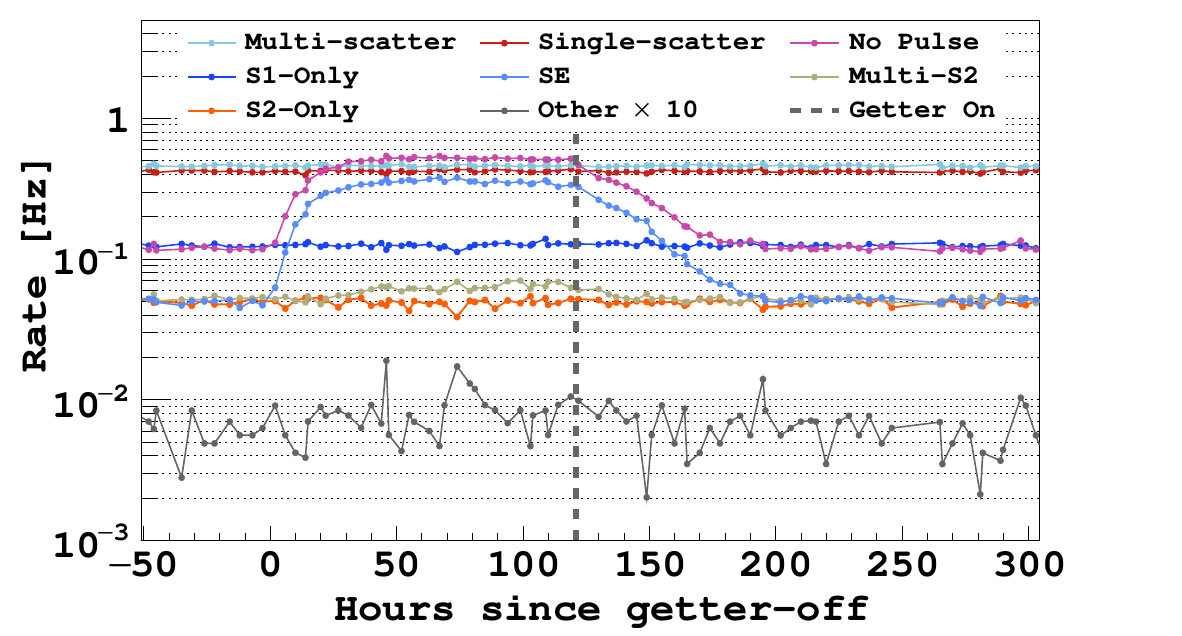}
 \caption{Rate of each event type in data collected with UAr. Top panel shows over the full dataset, and the bottom panel is zoomed in around day 100, showing the time the getter was bypassed (at \SI{0}{hours}) and back on (dashed gray line).}
 \label{fig:rate}
 \label{fig:rate_gettteroff}
\end{center}
\end{figure*}

\subsection{Time evolution of event rates}
The evolution of the event rate of each category, starting from April 1, 2015, the \UAr\ fill date, is presented in the top panel of Fig.~\ref{fig:rate}. 
Apart from ``SE'' and ``No Pulse'' events, each category remains steady over the \SI{1000}{days} of operation, reflecting the stability of the \TPC\ and cryogenic system. 
The gaps in time correspond to runs with different configurations, such as those with different field strengths or with calibration sources present. 
The slight decreasing trends over three years in ``Single scatter'' and ``Multi scatter'' events are consistent with the decay of \ce{^60Co} and \ce{^85Kr}.
For ``SE'' and ``No Pulse'' events, there are two decreasing trends: for the first \SI{200}{days}, the rate decreases with a \SI{\sim65}{day} decay constant, which then becomes \SI{\sim8}{years}.
There are also spikes in ``Other'' after roughly \SI{80}{days}, and in the ``SE'' and ``No pulse'' categories after roughly \SI{100}{days} from the \UAr\ fill date. 
The first was caused by an abnormally high rate in one \PMT\ and is not discussed further in this paper. 
The latter occurred when the getter was removed from the gas circulation system for maintenance for \SI{5}{days} starting on day 99.

\subsubsection{Getter-off period}


When the getter was bypassed in the gas circulation system, the \TPC\ event rate increased rapidly. 
The bottom panel of Fig.~\ref{fig:rate_gettteroff} shows the rate of each event category as a function of time since the removal of the getter. 
The rate increase is predominantly due to an increase in the ``No Pulse'' and  ``SE'' categories, composed of \STwo-like signals with $\Ne\num{<4}$. 
There is no corresponding increase in \STwo-only events with $\Ne\num{>4}$, demonstrating that such events behave differently than other \STwo-only events. 
The observed increase in ``No Pulse'' events is likely a result of misidentified SE events; due to the radial dependence of the \STwo\ yield, an SE-like event near the \TPC\ walls may be under-amplified relative to one near the center of the \TPC\ and fall below the pulse-finder's threshold. 
The \SI{\sim20}{\percent} increase in ``Multi-\STwo'' event rate (events with no \SOne-like signals but at least two \STwo-like signals) is likely due to SEs piling up with each other. The excess event rate increased for \SI{2}{days} and then stabilized until the getter was re-installed. After reinstallation, the excess event rate decreased with a \SI{36}{hour} time constant, which is in line with the purification rate of \SIrange{\sim20}{60}{hours} based on the circulation rate \SI{2.7}{kg/h}, an active \LAr\ mass of \SI{46}{\kg} in the TPC and a total \LAr\ mass of 155 kg in the cryostat.


\section{Evolution of purity metrics}
\label{sec:purity}
The spike in the SE rate when the getter is removed suggests a possible link to impurities in the \LAr. 
Two established measures of the \LAr\ purity are the free electron lifetime $\tau_e$, which measures the fraction of electrons captured by electronegative impurities before reaching the gas pocket, and the lifetime of the long-lived triplet component in \SOne, $\tau_\ell$, which measures the abundance of impurities that can non-radiatively dissipate energy stored in argon excimers, such as through vibrational degrees of freedom.
Although these metrics are not sensitive to all possible impurity types and may have limited sensitivity to dilute impurities, they can still constrain impurities potentially linked to SE production.

\subsection{Free electron lifetime}
Since electronegative impurities such as \ce{O_2} may capture drifting electrons~\cite{Bakale1976EffectOA,chaninMeasurementsAttachmentLowEnergy1962a}, electrons from ionization events closer to the cathode are more likely to be captured before reaching the gas pocket than those produced near the grid. 
The free-electron lifetime quantifies the average time in which electrons drift before being captured. 
This is a function of the electronegative impurity concentration and electron drift speed.
To evaluate $\tau_e$, the dependence of \STwo\ charge on vertical distance between the interaction site and cathode is measured, using both physics data and a mono-energetic \ce{^{83m}Kr} calibration source with a \SI{1.83}{hour} half-life injected in the \UAr.

\subsubsection{Measurements with physics data}
\begin{figure}[htb]
\begin{center}
\includegraphics[width=\columnwidth]{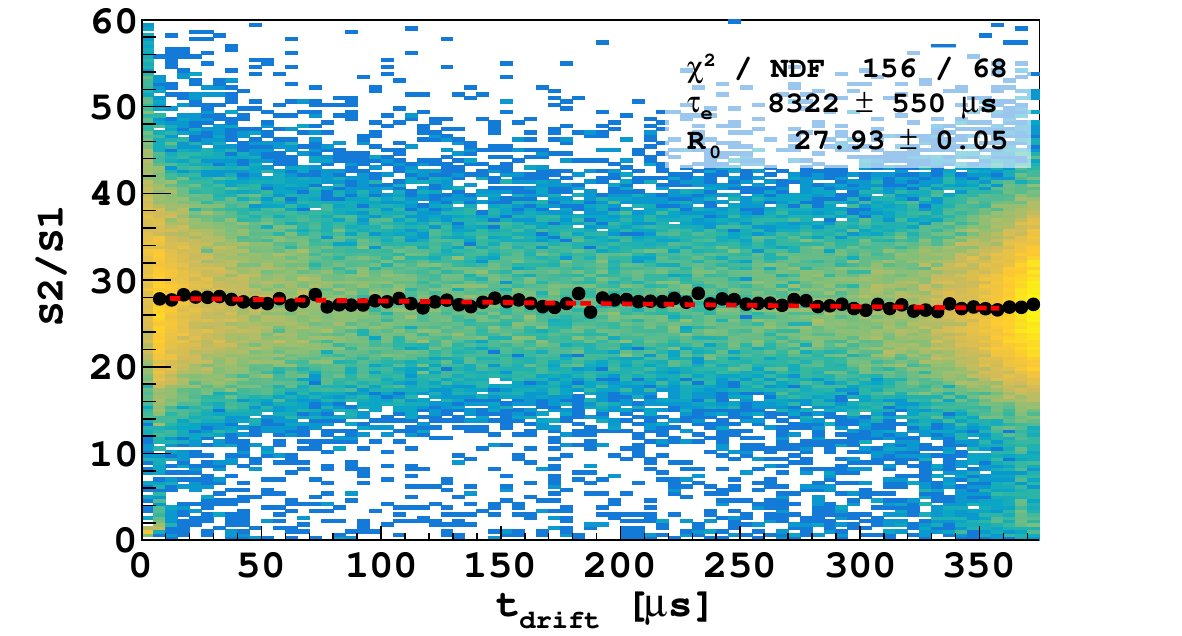}
 \caption{Dependence of S2/S1 on \tdrift\ with the getter off. The black points are mean values in each \tdrift\ bin and the red dashed line is the exponential fit function.}
 \label{fig:s2/s1_gettteroff}
\end{center}
\end{figure}

During physics data-taking, with no calibration sources present, background \gr\ peaks are available for free electron lifetime determination.
The value of $\tau_e$  is estimated from the relationship between drift time (\tdrift)\  and the ratio \STwo/\SOne, after correcting \STwo\ for horizontal position and \SOne\ for vertical position. 
After removing the observed energy dependency of \STwo/\SOne, the mean value of \STwo/\SOne\ at a given \tdrift\ is proportional to the fraction of ionization electrons that reach the gas pocket.
The observed relationship between \STwo/\SOne\ and \tdrift\ is shown in Fig.~\ref{fig:s2/s1_gettteroff} for the runs during the getter off period as an example.
A function $f(\tdrift)={\rm R}_0 e^{-\tdrift/\tau_e}$ is fit to the mean value of \STwo/\SOne, where ${\rm R}_0$ and $\tau_e$ are fit parameters. The best value of $\tau_e$ obtained in Fig.~\ref{fig:s2/s1_gettteroff} is \SI{8.3}{ms} for the getter-off runs.

\subsubsection{Measurements with \ce{^{83m}Kr} calibration data}
A \ce{^{83m}Kr} calibration source was injected into the \UAr\ approximately annually to study $\tau_e$ and calibrate the \SOne\ light yield. 
\ce{^{83m}Kr} produces mono-energetic signals via sequential decays of \SI{32.1}{keV} and \SI{9.4}{keV}, separated with a half-life of \SI{154}{ns}. 
To determine the electron lifetime, \ce{^{83m}Kr} events are selected around their \SOne\ peak, and mean \STwo\ charge as a function of \tdrift\ is examined, after subtracting the background spectra measured in physics runs as necessary. 
The mean \STwo\ is then fit as a function of \tdrift.

\begin{figure}[htb]
\begin{center}
\includegraphics[width=\columnwidth]{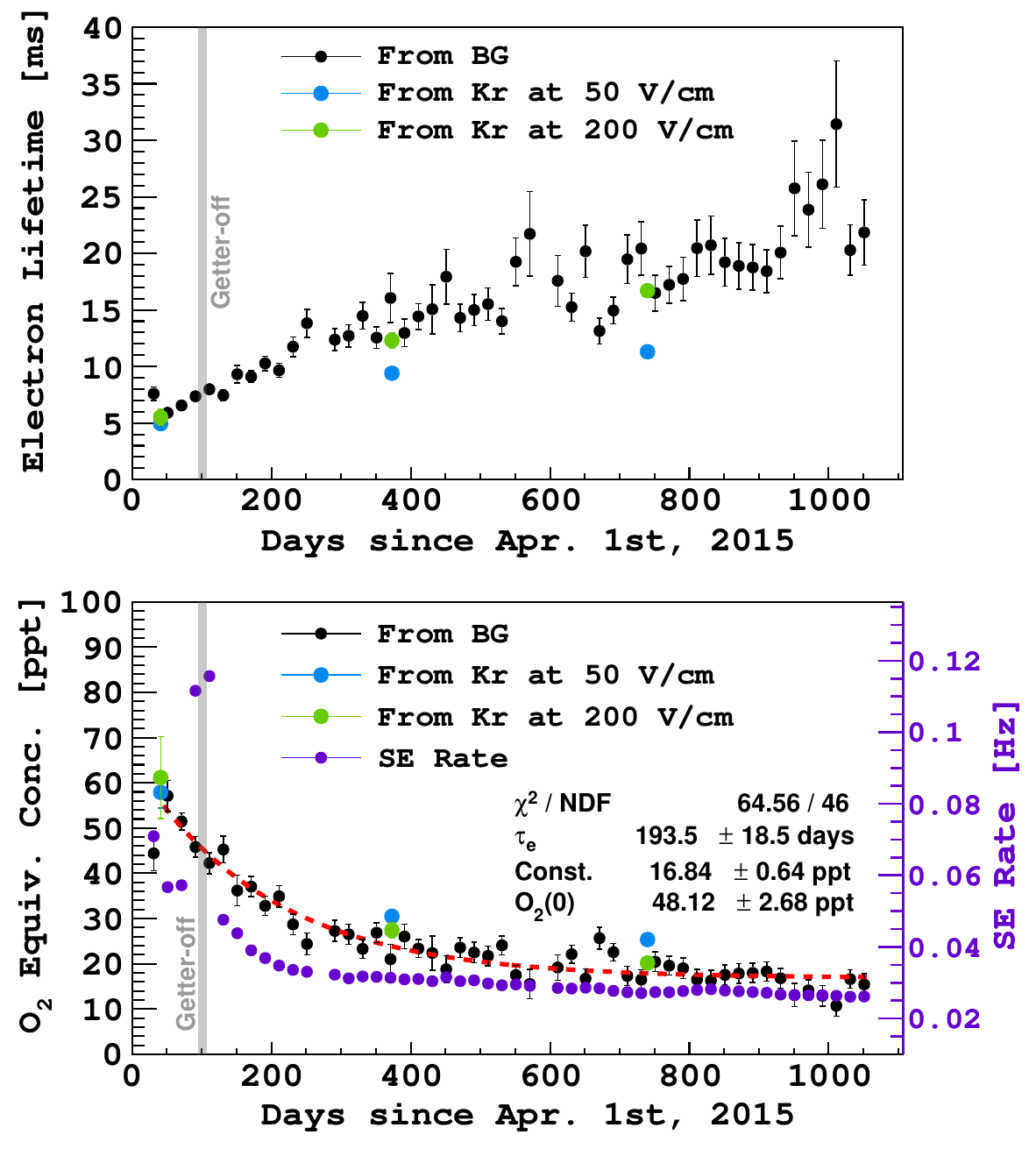}
 \caption{(Top) Electron lifetime vs time (Bottom) \ce{O_2} equivalent contamination vs time with time evolution of the total SE rate. The red dashed line shows an exponential function fit to the \ce{O2} equivalent concentration inferred from background (BG) data. The Getter-off days are indicated in gray. }
 \label{fig:electron_lifetime}
\end{center}
\end{figure}

The top panel in Fig.~\ref{fig:electron_lifetime} shows $\tau_e$ measured in physics and calibration data.
For $\tau_e\SI{>10}{\ms}$, the decrease in \STwo\ at large \tdrift\ becomes small, resulting in large statistical uncertainties with a \SI{200}{\volt\per\cm} drift field.
To decrease these uncertainties, $\tau_e$ is also measured at a \SI{50}{\volt\per\cm} drift field with the \ce{^{83m}Kr} calibration source, where the slower drift speed and higher electron attachment coefficients result in more electron loss~\cite{Bakale1976EffectOA,liParameterizationElectronAttachment2022}.

When the getter was removed, $\tau_e$ did not decrease: the measured value of \SI{8.3}{\ms} is consistent before and after this period. 
The impurities that dominate the electron lifetime therefore cannot alone explain the rapid increase in the SE rate when the getter was turned off.

Taking \ce{O2} as a representative electronegative impurity, the \ce{O2} concentration $[\ce{O2}]$ consistent with $\tau_e$ is shown in the bottom panel of Fig.~\ref{fig:electron_lifetime}, related by~\cite{Acciarri:2010jh},
\begin{align*}
    [\ce{O_2}]/[\ce{Ar}] = 1/(35 k\tau_e),
\end{align*}
for \ce{Ar} concentration $[\ce{Ar}]=\SI{35}{\mole\per\liter}$ and rate constant $k=\SI{9e10}{\liter\per\mole\per\second}$ at \SI{200}{\volt\per\cm} and $\SI{e11}{\liter\per\mole\per\second}$ at \SI{50}{\volt\per\cm} ~\cite{Bakale1976EffectOA,liParameterizationElectronAttachment2022}. 
Fitting an exponential distribution to the change in 
$[\ce{O2}]$ over time gives a decay constant of $\sim$\odecay. 
There is no clear correlation between the impurities responsible for the finite electron lifetime and the SE event rate, suggesting that the impurities that dominate electron lifetime are not primarily responsible for SEs. 
Similar results have been found for LXeTPCs, including LUX and XENON1T~\cite{akeribInvestigationBackgroundElectron2020,Aprile_2021}.


\subsection{Pulse shape}
\label{sec:pulseshape}

\SOne\ pulse shape fits in \LAr\ were performed by WArP~\cite{WArP:2008rgv}, DEAP-3600~\cite{DEAP:2020hms}, DUNE~\cite{DUNE:2022ctp}, and ARIS~\cite{Agnes_2021b}. 
%
WArP studied the pulse-shape dependence on \ce{N2} concentration $[\ce{N2}]$, showing changes in the triplet component's amplitude $p_\ell$ and lifetime $\tau_\ell$ for $[\ce{N2}]\SI{>1}{\ppm}$. 
This effect is understood to be due to quenching from non-radiative reactions between \ce{Ar2} dimers and \ce{N2} molecules. 
Since triplet dimers live longer than singlets, they are more heavily suppressed.
For $[\ce{N2}]\gtrsim\SI{1}{\ppm}$, it can be estimated using $\tau_\ell$~\cite{WArP:2008rgv}.
%
%
The impact of \ce{N2} in \DSf\ is measured by searching for correlations between $\tau_\ell$ and the SE rate.

To estimate $\tau_\ell$, \num{50000} waveforms are averaged in \SI{2}{day} intervals, requiring:
(1) $\SOne\in[\numrange[range-phrase=\ensuremath{,{}},range-units=single]{100}{20000}]\,\si{\pe}$;
(2) exactly \num{2} pulses were identified;
(3) \tdrift$>$\SI{20}{\us}; 
(4) \SI{\geq20}{\ms} have passed since the previous event;
(5) $\num{0.15}<f_{90}<\num{0.5}$;
(6) and top-bottom asymmetry $A_\text{t-b}\in[\numrange[range-phrase=\ensuremath{,{}}]{-0.9}{0.9}]$, where \mbox{$A_\text{t-b}=(\text{PE}_\text{t}-\text{PE}_\text{b})/(\text{PE}_\text{t}+\text{PE}_\text{b})$} and $\text{PE}_\text{t(b)}$ is the number of PE in the top (bottom) \PMT\ array.
These criteria select isolated, single-scatter electronic recoils from the bulk \LAr, at least \SI{2}{\cm} below the extraction grid.

\begin{figure}[htb]
\begin{center}
\includegraphics[width=\columnwidth]{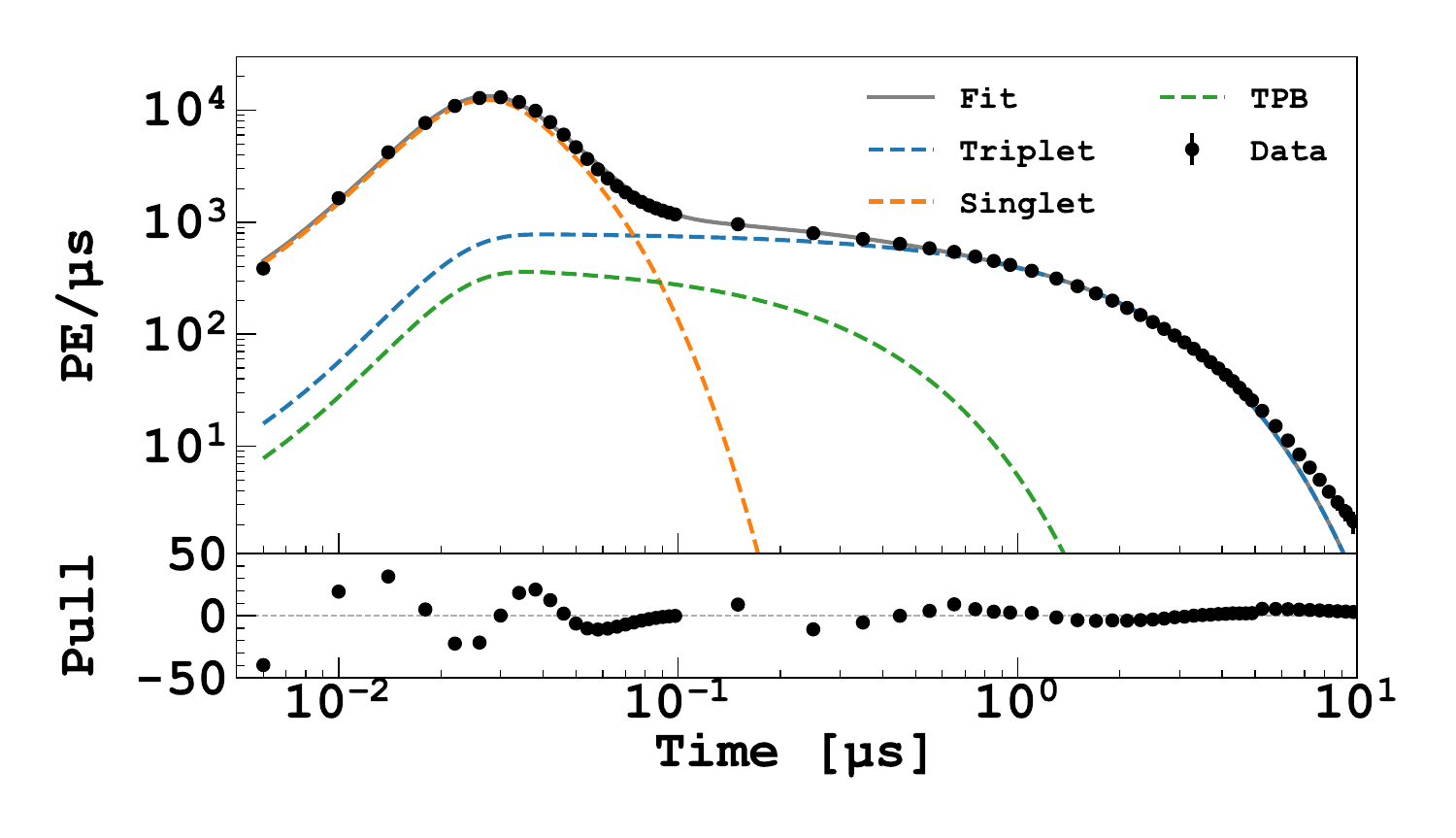}
 \caption{Example average waveform. 
(Top) Fit with $F(t)$ from Eq.~\ref{eq:pulseshape} (gray), with contributions from the singlet (orange), triplet (blue), and \TPB\ (green) components shown. (Bottom) Pulls from the fit, (Data-Func.)/Error of Data. }
 \label{fig:pulse_time_1}
\end{center}
\end{figure}

The average waveform is fit using the function $F$ described in Sec.~3.1 of Ref.~\cite{Agnes_2021b}; 
\begin{gather}
    \begin{aligned}
        &F(t|p_\text{TPB},p_s,p_\ell,\tau_s,\tau_\ell,\sigma,t_0,A,C) = C + A\times\\
        &\sum_{i=\{s,\ell\}}p_i\left[(1-p_{_\text{TPB}}) P_0(t^\prime|\tau_i,\sigma)+p_{_\text{TPB}} P_1(t^\prime|\tau_i,\tau_{_\text{TPB}},\sigma)\right],\\
        &\text{where }t^\prime=t-t_0,\\
        &P_0(t^\prime|\tau_i,\sigma) = \frac{1}{2\tau_i}\left(1+\text{erf}\left(\frac{t^\prime-\sigma^2/\tau_i}{\sqrt{2}\sigma}\right)\right)e^{\frac{-t^\prime}{\tau_i}+\frac{\sigma^2}{2\tau_i^2}} \text{,} \\
        &P_1(t^\prime|\tau_i,\tau_\text{TPB},\sigma)=\frac{\tau_iP_0(t^\prime|\tau_i,\sigma)-\tau_\text{TPB}P_0(t^\prime|\tau_\text{TPB},\sigma)}{\tau_i-\tau_\text{TPB}}.
        \label{eq:pulseshape}
    \end{aligned}
\end{gather}
This model describes two exponential decays with probabilities $p_s$ and $p_\ell$ and lifetimes $\tau_s$ and $\tau_\ell$ corresponding to the short-lived (singlet) and long-lived (triplet) components.
This distribution is convolved with the \TPB's re-emission modeled with a prompt and a delayed component, and a Gaussian distribution with timing offset $t_0$ and resolution $\sigma$. 
The present analysis differs from Ref.~\cite{Agnes_2021b} in two ways: (1) only one TPB component is considered, and (2) a constant term $C$ is fixed to \num{0}.
Both differences have negligible effects on the present study.
Figure~\ref{fig:pulse_time_1} shows one such fit, performed from the start of the waveform to \SI{10}{\micro\second}. 

Based on the fit results, no correlation is observed between $p_\ell$ or $\tau_\ell$ with the SE rate, including the getter-off period. 
At the end of the data set, where the electron lifetime is longest, the triplet lifetime $\tau_\ell$ is \SI{1.377\pm0.008}{\micro\second}, which is consistent with Refs.~\cite{DEAP:2020hms,DUNE:2022ctp}.
Throughout the data collection period, $\tau_\ell$ decreases by $<\SI{0.050}{\micro\second}$ (\SI{90}{\percent} C.L.).
Allowing $C$ to be nonzero in these fits increases $\tau_\ell$ by \SI{4}{\percent}; also accounting for the second \TPB-re-emission component increases $\tau_\ell$ by \SI{14}{\percent}. 
Varying the fit range from \SIrange{5}{15}{\micro\second} increases $\tau_\ell$ by up to \SI{3}{\percent}. 
No energy-dependence of $\tau_\ell$ is observed within the considered PE range.


Based on these results and those in  Ref.~\cite{WArP:2008rgv}, the increase of SE rate during the getter-off period is not linked to increased \ce{N2} contamination above the \SI{1}{\ppm} level.


\section{Observed spurious electron correlations}
\label{sec:correlations}
Several mechanisms have previously been proposed to explain the SEs observed in liquid xenon \TPCs, including chemical impurity interactions, delayed electron extraction into the gas phase~\cite{zeplin11}, spontaneous field emission of the grid, and de-excitation of metastable target excimers~\cite{Sorensen:2017kpl,zeplin11,Aprile_2014,akeribInvestigationBackgroundElectron2020}.
Many of these mechanisms may also be applicable to \LAr\ \TPCs\ and can be studied by examining the correlations between SEs and preceding events.


A subset of ``Single-scatter'' and ``Multi-scatter'' events that may be followed by SEs, called ``parent events" (or simply "parents" in the following), are selected for the study of SE event correlations.
Unless otherwise specified, they are required to have $\SOne\SI{>1000}{\pe}$ after correcting for the vertical position, one \SOne, at least one \STwo, and a successfully-reconstructed horizontal position. 
This \SOne\ threshold is chosen to increase the probability of the parent being followed by SEs, as will be discussed later.

\subsection{Time correlation with previous events} \label{sec:time_corr}

\begin{figure}[htb]
\begin{center}

\includegraphics[width=\columnwidth]{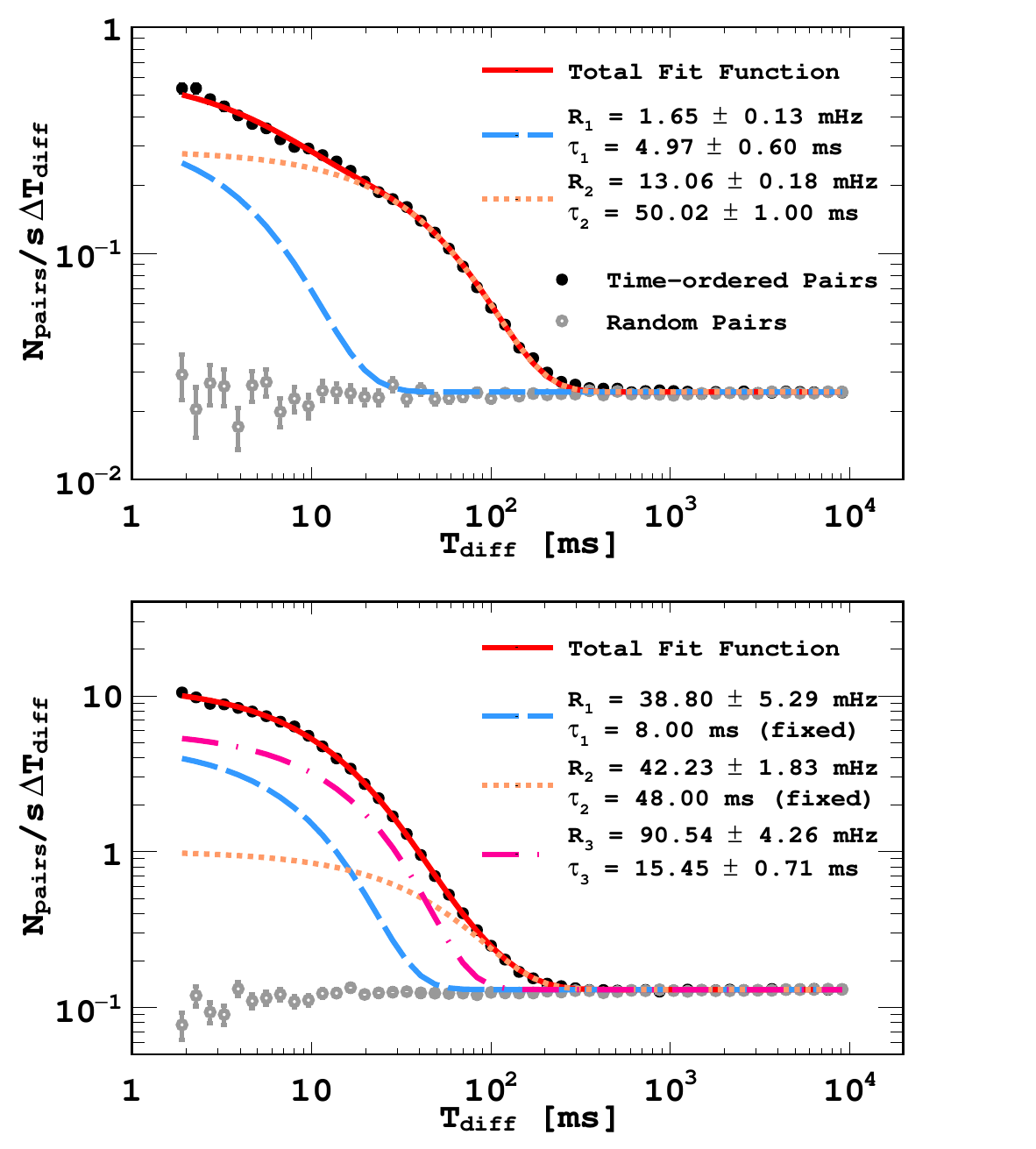}
 \caption{
 The time difference $T_{diff}$ separating SEs and all preceding parent events in a \SI{10}{\s} window, normalized to lifetime and bin width.
 Closed black circles show SEs following parents (time-ordered pairs), while open gray circles show random pairs, which have a flat distribution as expected.
 The red solid line shows the total fit function, with two exponential components and a constant. (Top) Shows physics data from runs \SIrange[range-units=single,range-phrase=\mbox{--}]{319}{353}{days} after the UAr fill, fitted with two exponential components (dashed blue and dotted orange lines) and a constant. (Bottom) Shows data with the getter off, fitted with two exponential component with fixed $\tau$s from fit just before the getter off period, an
additional free exponential component (dash-dotted pink line), and a constant. $\Delta T_{diff}$ is the bin width of x-axis.
 }
 \label{fig:tcorr}
 \label{fig:tcorr_gettteroff}
\end{center}
\end{figure}

\begin{figure*}[htb]
\begin{center}
\includegraphics[width=\textwidth]{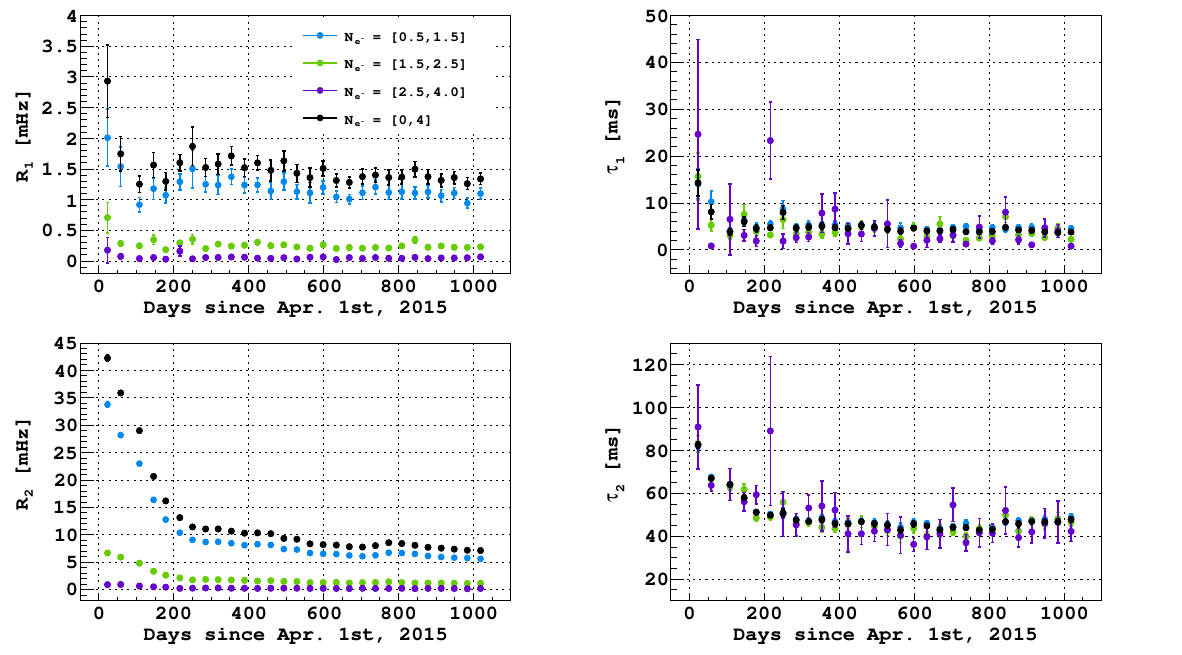}
 \caption{$T_{diff}$ fit parameters for various \Ne\ ranges, excluding runs for which the getter was turned off.}
 \label{fig:tcorr_param}
\end{center}
\end{figure*}

To study coincidences between SEs and parents, the time difference $T_{diff}$ separating each SE from all preceding parent candidates in a \SI{10}{\second} window (time-ordered pairs) is calculated. The contribution of random coincidences is estimated using $T_{diff}$ for parents that follow SEs in the same time span (random pairs). Values of $T_{diff}$ are binned in a histogram with logarithmic bin widths, reflecting the correlation's exponential nature.
Due to the inhibit time and veto windows used for this analysis, these fits are not sensitive to correlations on \SI{<1}{\milli\second} timescales.
Much shorter correlations, with \SI{<370}{\micro\second} delays, are discussed in Ref.~\cite{agnesStudyEventsPhotoelectric2022a}, which explores effects such as photoelectric emissions.

Runs were grouped in \SI{35}{day} intervals, excluding runs with the getter off. The average rate of SE-parent pairs in each $T_{diff}$ bin is calculated by dividing the number of pairs by the livetime of the \SI{35}{day} interval. The resulting $T_{diff}$ distribution, after normalizing by the width of each $T_{diff}$ bin ($\Delta T_{diff}$), is well-described by the sum of two exponential functions:
\begin{equation}
    f(T_{diff}) = \frac{R_1}{\tau_1}e^{-T_{diff}/\tau_1}+\frac{R_2}{\tau_2}e^{-T_{diff}/\tau_2} + C,
    \label{eq:tcorr_fitfunc}
\end{equation}
where $R_1$ and $R_2$ are the rates of two components with decay constants $\tau_1$ and $\tau_2$, with $\tau_1 < \tau_2$, and $C$ accounts for random coincidences. 
For data with the getter off, an additional component with an intermediate decay constant is needed to describe the data. 

The top panel of Fig.~\ref{fig:tcorr} shows the $T_{diff}$ distribution observed over a \SI{35}{day} period starting \SI{319}{days} after the UAr fill.
Equation~\ref{eq:tcorr_fitfunc} is fit to this distribution, showing strong agreement between data and this decay model.
With the getter off, a component with a \SI{15}{\ms} decay constant appears, as shown in the bottom panel of Fig.~\ref{fig:tcorr_gettteroff}, with $\tau_1$ and $\tau_2$ fixed with their values before the getter-off period.
Reduced $\chi^2$ values for these fits are between \numrange{0.9}{1.7}. 

Figure~\ref{fig:tcorr_param} shows the results of the fits to $T_{diff}$ distributions, performed for 35-day ranges during the $\sim$1000-day data collection period.  
During each interval, events were analyzed separately according to their \Ne\ values. These fits show that $R_1$ and $\tau_1$ are stable near \SI{1.5}{\milli\hertz} and \SI{5}{\ms}, respectively.
However, $R_2$ and $\tau_2$ decrease significantly during the first \SI{200}{days} and continuously decrease afterwards at a slower rate: $\tau_2$ decreases from \SI{80}{\milli\second} to \SI{40}{\milli\second}, and $R_2$ decreases from \SI{43}{\milli\hertz} to \SI{10}{\milli\hertz}. 

Temporal variations in $\tau_1$ and $\tau_2$ are independent of \Ne, potentially indicating that SEs with different \Ne\ share a common origin.

\begin{figure}[htb]
\begin{center}
\includegraphics[width=\columnwidth]{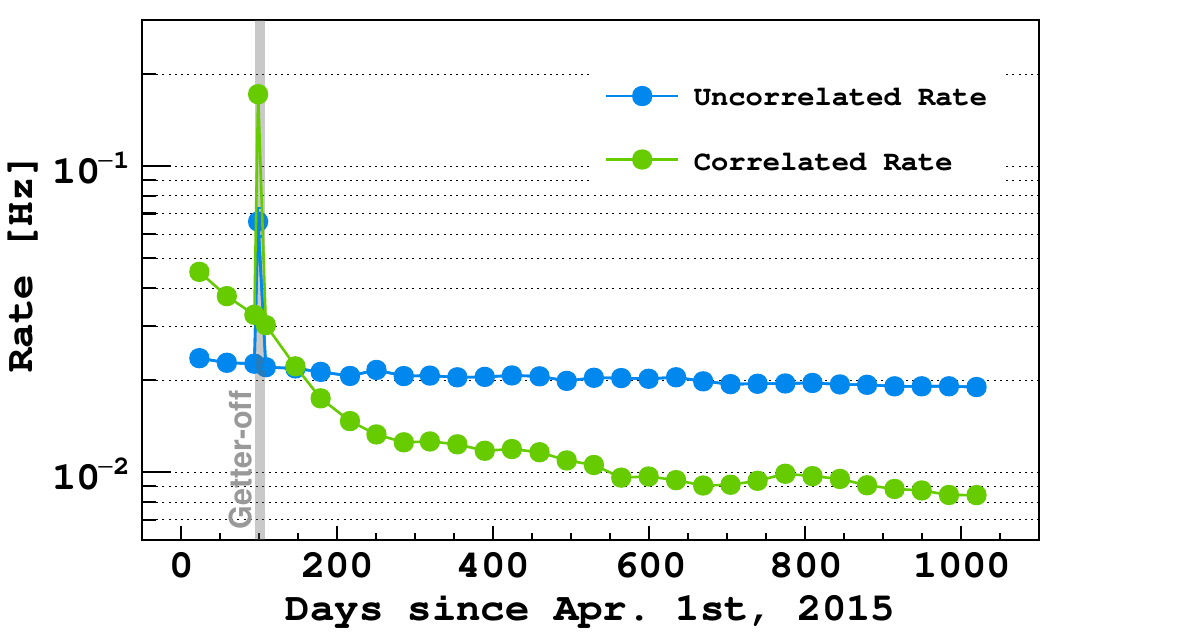}
 \caption{Rate of correlated SE events (green), found by summing the $R_1$ and $R_2$ fit parameters. Rate of remaining uncorrelated events is plotted in blue. Days with the Getter-off are indicated in gray.}
 \label{fig:corr_uncorr_rate}
\end{center}
\end{figure}

The sum $R_1+R_2$ gives the total rate of SEs with an identified temporally-correlated parent. Note that this temporally correlated rate may be slightly underestimated due to missing parents that fall outside the parent selections or within the inhibit-time window.
Figure~\ref{fig:corr_uncorr_rate} compares the correlated SE rate to the uncorrelated SE rate, defined as the remainder of the total SE rate from Fig.~\ref{fig:rate} after subtracting $R_1+R_2$, consistent with SEs with no identified parent. 
While the correlated SE rate decreases significantly in the first \SI{200}{days} and then more slowly, the uncorrelated rate remains approximately constant while the getter was in use.
When the getter was bypassed, the rate of correlated SEs increased 6-fold, while the uncorrelated SE rate tripled. 
At the start of data collection, the correlated component accounts for approximately \SI{70}{\percent} of all SEs,  decreasing to \SI{30}{\percent} after \SI{200}{days}.

Given the low probability of seeing SEs from low-energy parents (as will be discussed in \refsec{energy_corr}), changing the parent threshold from $\SOne>\SI{1000}{\pe}$ does not significantly change the fraction of correlated events. 
Specifically, lowering the parents-selection threshold to $\SOne>\SI{50}{\pe}$ increases the correlated component fraction by only 2\%.

\begin{figure}[htb]
\begin{center}
\includegraphics[width=\columnwidth]{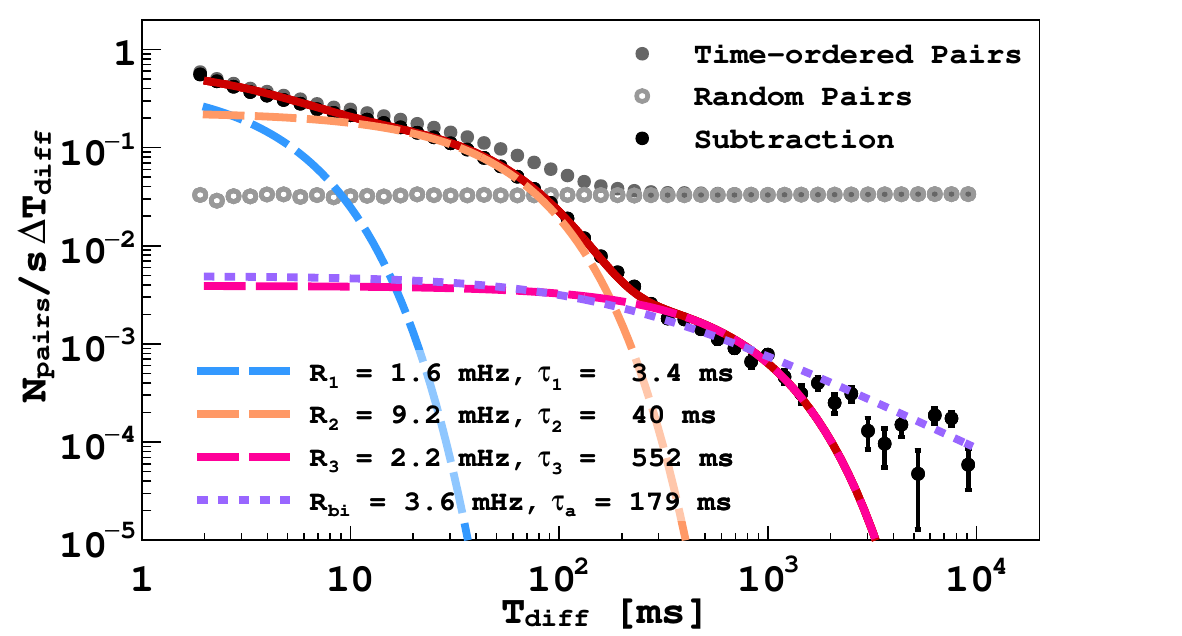}
 \caption{The time difference $T_{diff}$ observed over \SI{500}{days}. The black solid points are the difference of time-ordered pairs from random pairs with a three-exponential function fit in a red solid line. 
The blue, orange, and pink dashed lines show the three exponential components, respectively. The purple dotted line shows the biexcitonic ionization function. $\Delta T_{diff}$ is the bin width of x-axis.}
 \label{fig:dRdT_long}
\end{center}
\end{figure}

Figure~\ref{fig:dRdT_long} shows the same analysis with runs grouped into longer time intervals, to increase statistics at larger values of $T_{diff}$.
The observed $T_{diff}$ distribution is selected from a late \SI{500}{day}-period where $\tau_{1,2}$ and $R_{1,2}$ are stable. 
The data suggest the existence of a subdominant long-lived component not described by the two-exponential model. 
A fit with the sum of three exponential functions (a simple extension of Eq.~\ref{eq:tcorr_fitfunc}) returns a decay constant around \SI{552\pm33}{\milli\second} for the third exponential function, constituting \SI{\sim7}{\percent} of all SEs. Figure~\ref{fig:dRdT_long} suggests that a longer-lived component ($T_{diff} > 1 s$) may behave like a power law at late times but is subdominant at earlier times. 
This behavior may also be a result of exponential contributions that vary over time, obfuscated by low statistics.
However, as shown in Fig.~\ref{fig:dRdT_long}, the $T_{diff}$ distribution is also well-described by a third component that evolves as $dR_{bi}/dt = N_{bi}/(1+t/t_a)$, where $N_{bi}$ and $t_a$ are constants. 
This function can describe the decay rate of a long-lived state via biexcitonic ionization processes, which decrease with a time scale $t_a$ due to diffusive processes, and is consistent with the functional form derived in Ref.~\cite{voltzRadioluminescenceMilieuxOrganiques1968,hitachiLuminescenceQuenchingLiquid1992} and explored in Ref.~\cite{PhysRevD.98.062002} in the limit where the diffusing species are long-lived and sparse. This function was fit to the data along with two exponential functions by fixing their parameters to the result of the three-exponential function fit.
The best-fit parameters for $dR_{bi}/dt$ allow this component to describe about \SI{12}{\percent} of all SEs.

\STwo-only events with $\Ne>4$ do not show temporal correlations with preceding events in a \SI{1}{\second} window, indicating that the observed correlation is particular to SEs.

\subsection{Correlation with total ionization activity} \label{sec:long_lived}

As shown in Fig.~\ref{fig:corr_uncorr_rate}, about half of SEs are described by correlated components $R_1$ and $R_2$, which vary over time.
This method of identifying correlations loses sensitivity for correlation timescales that are long compared to the total \SI{1.5}{\hertz} event rate.
SE production mechanisms related to parents may therefore not be counted as ``temporally correlated'' if their delay is too long.

\begin{figure}[htb]
\begin{center}
\includegraphics[width=\columnwidth]{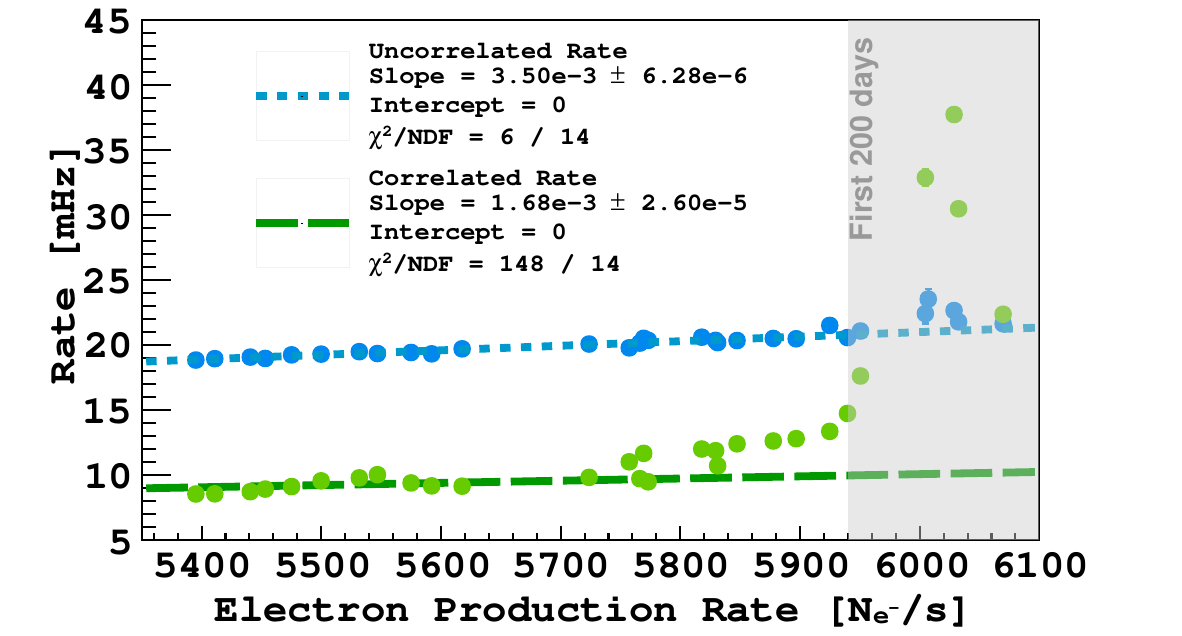}
 \caption{The rate of correlated (green) and uncorrelated SEs (blue) as a function of total electron production rate in \SI{35}{day} intervals excluding getter-off runs. Linear fits are performed on runs collected after the first \SI{200}{days}. The shaded area indicates data from the first 200 days.}
 \label{fig:rate_v_Ne}
\end{center}
\end{figure}

Figure~\ref{fig:rate_v_Ne} shows the SE rate averaged over \SI{35}{day} intervals as a function of the average ionization rate, measured as the total \Ne\ in all non-SE \STwo\ pulses.
The correlated SE rate from a single impurity is proportional to the ionization rate; however, since SEs have multiple sources that vary in intensity (\textit{e.g.}, impurities that are removed over time), the final relationship is not strictly linear.
Since uncorrelated SEs may not be related to energy depositions in the \LAr, such proportionality is not guaranteed---a constant SE production rate may be present in the limit where no ionization events occur in the \LAr. 
This relationship is explored in Fig.~\ref{fig:rate_v_Ne} by a linear fit with some $y$-intercept corresponding to the SE rate expected with no ionization within the \TPC, and a slope describing SEs related to ionization events.

After the first \SI{200}{days}, this trend is well-described by a linear fit with the intercept fixed to 0, implying that the data are consistent with all apparently-uncorrelated SEs actually having an undetected parent. 
Varying the intercept value and repeating the fit show that uncorrelated SEs have a rate \SI{<7.5}{\milli\hertz} (\SI{90}{\percent} C.~L.). The uncorrelated rate in Fig.~\ref{fig:corr_uncorr_rate} is approximately \SI{20}{\milli\hertz}, meaning that less than a third of this component can be independent of total TPC ionization activity.

\subsection{Energy correlation with previous events} \label{sec:energy_corr}

To study the energy dependence of SEs, they are paired with the most recent single-scatter parent candidate, relaxing the parent identification threshold to $\SOne\SI{>50}{\pe}$. 
Given that the rate of parent events is \SI{\sim1.5}{\hertz} and $\tau_2\SI{\sim50}{\milli\second}$, this procedure is robust in identifying the correct SE parent. 
Based on the fit to the SE timing distribution in Fig.~\ref{fig:dRdT_long}, the fraction of mismatched parents is estimated to be at most $\sim 0.5\%$.

%

\begin{figure}[htb]
\begin{center}
\includegraphics[width=\columnwidth]{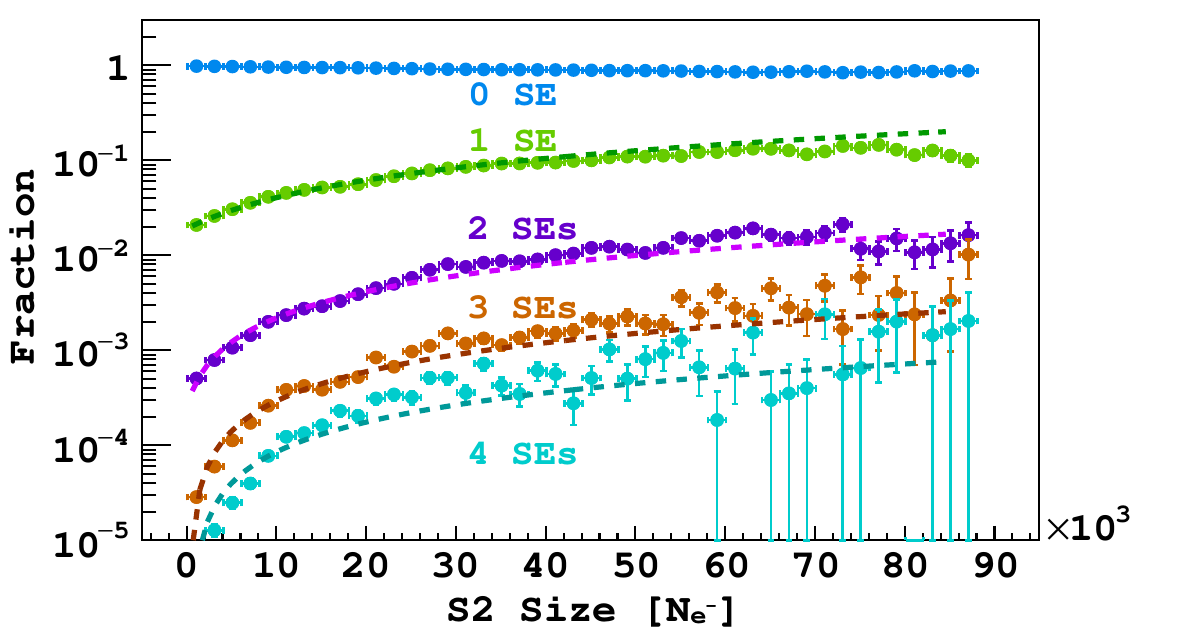}
 \caption{Fractions of parent events followed by different numbers of SE events as a function of the S2 size of the parent. The dashed lines show linear fits. }
 \label{fig:S2_Parent_Fraction}
\end{center}
\end{figure}
Figure~\ref{fig:S2_Parent_Fraction} shows the fraction of parents followed by \numrange[range-phrase=--]{0}{4} SE events before the occurrence of the next parent as a function of its \STwo\ size.
The number of SEs following a parent event increases linearly with the parent's \STwo, consistent with the hypothesis that SEs are produced by drifting ionization electrons. 

\begin{figure}[htb]
\begin{center}
\includegraphics[width=\columnwidth]{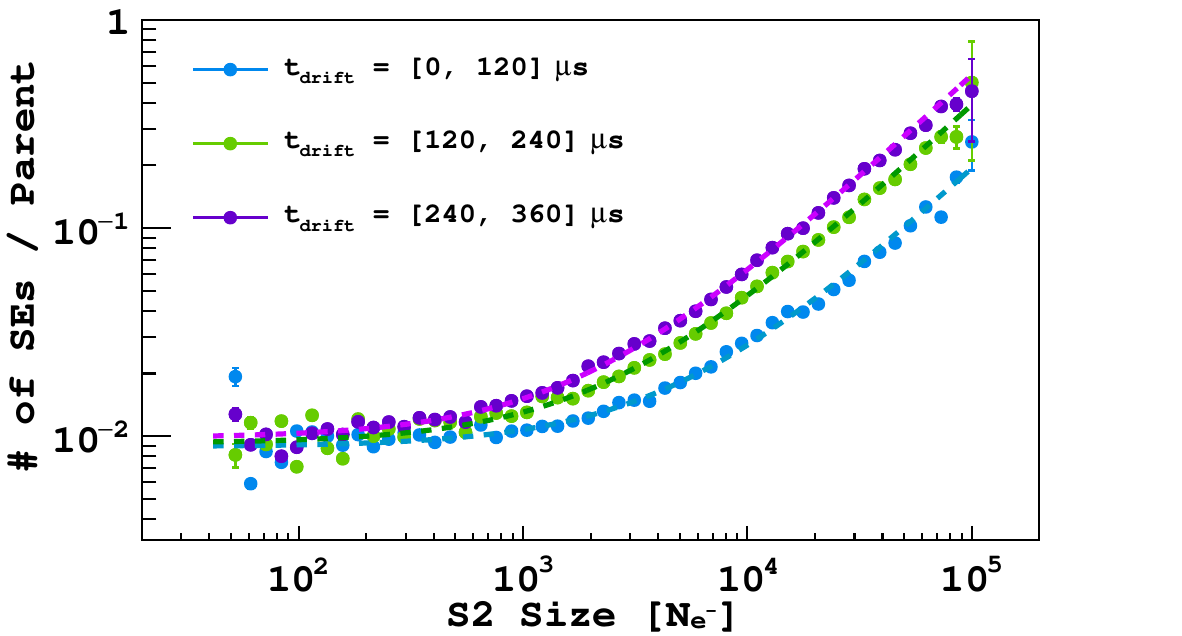}
 \caption{Number of SE events following parent events divided by the number of parent events with a given S2 size. Colors represent cuts in \tdrift\ of parent events. Dashed lines show linear fits.
 }
 \label{fig:SEvS2}
\end{center}
\end{figure}

Figure~\ref{fig:SEvS2} shows the average number of SEs following a parent event, the ratio of the number of SE events to parent events, as a function of parent \STwo\ charge $\STwo_\text{parent}$. 
If SEs are produced with some constant probability per parent electron, this ratio should approach zero for no $\STwo_\text{parent}$, and increase linearly with $\STwo_\text{parent}$. However, for low-energy parent events, the SE probability approaches a constant value, corresponding to a component of SEs seemingly uncorrelated with parent energy. 
This component may represent parent-SE pairs where the given SE event originated from some parent other than the one immediately preceding it, caused by a long-lived correlated component (as described in \refsec{long_lived}).
The expectation value of the number of SEs following a parent event $p_{SE}$ can be modeled with
\begin{equation*} \label{eq:S2_size_fit}
    p_{SE}(\STwo_\text{parent}) = c_1 + c_2 \cdot {\STwo_\text{parent}},
\end{equation*}
where $c_1$ and $c_2 \cdot \STwo_\text{parent}$ describe uncorrelated and correlated SEs, respectively. 
Best-fit values of $c_1$ and $c_2$ are shown in Table~\ref{tab:SEvS2}. 
At low energies, SEs are dominated by the uncorrelated component, produced at a constant rate.
As $\STwo_\text{parent}$ increases, correlated SEs become more likely, and $p_{SE}$ scales linearly with $\STwo_\text{parent}$. 

\begin{table}[htb]
\caption{Fit parameters to SE event probability as a function of parent S2 size, as plotted in Fig.~\ref{fig:SEvS2}. }
\label{tab:SEvS2}
\centering
\begin{tabular}{c|c|c}\hline\hline
$t_\text{drift}$\,[\si{\micro\second}] & $c_1$ & $c_2$ [$1/\Ne$]  \\ \hline
{[}0, 120{]} & $(8.97 \pm 0.13)\times10^{-3}$ & $(1.85 \pm 0.03)\times10^{-6}$ \\
{[}120, 240{]} & $(9.23 \pm 0.19)\times10^{-3}$ & $(3.82 \pm 0.06)\times10^{-6}$ \\
{[}240, 360{]} & $(9.82 \pm 0.14)\times10^{-3}$ & $(5.34 \pm 0.05)\times10^{-6}$
\\\hline\hline
\end{tabular}
\end{table}

Figure~\ref{fig:SEvS2} also shows that the SE event probability depends on $\STwo_\text{parent}$ for different ranges of $t_\text{drift}$ of parent events. 
For low-energy parents, with $\STwo\lesssim\SI{e3}{\el}$, the SE event probability varies little with \tdrift: $c_1$ increases by \SI{10}{\percent} as \tdrift\ increases from the lowest to highest interval shown in Table~\ref{tab:SEvS2}. 
As correlated SEs become more prominent at higher $\STwo_\text{parent}$, the SE probability increases with \tdrift, as $c_2$ increases by a factor of \num{2.9}.

\subsection{Drift time correlations with previous events}
In order to study spatial correlations between SE and parent events, parent event selections are tightened to ensure that event positions are well-defined by requiring parent events to be single scatter events (one \SOne\ and one \STwo).
\begin{figure}[htb]
\begin{center}
\includegraphics[width=\columnwidth]{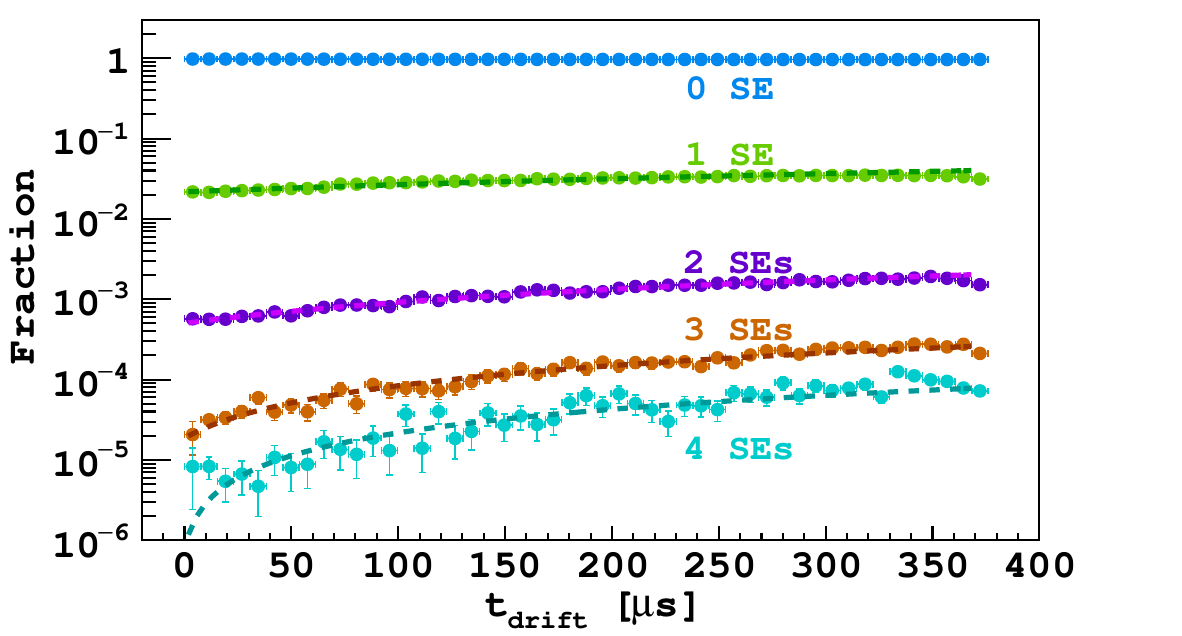}
 \caption{Fractions of parent events followed by different numbers of SE events as a function of parent \tdrift. The dashed lines show linear fits. }
 \label{fig:Tdrift_Parent_Fraction}
\end{center}
\end{figure}

To examine the relationship between parent \tdrift\ and SE probability, the fraction of parents followed by an SE is measured as a function of \tdrift, summing all SEs that follow the parent until the occurrence of the next parent.
Figure~\ref{fig:Tdrift_Parent_Fraction} shows that the expected number of SEs increases with longer parent \tdrift, as electrons drift greater distances.

\begin{figure}[htb]
\begin{center}
\includegraphics[width=\columnwidth]{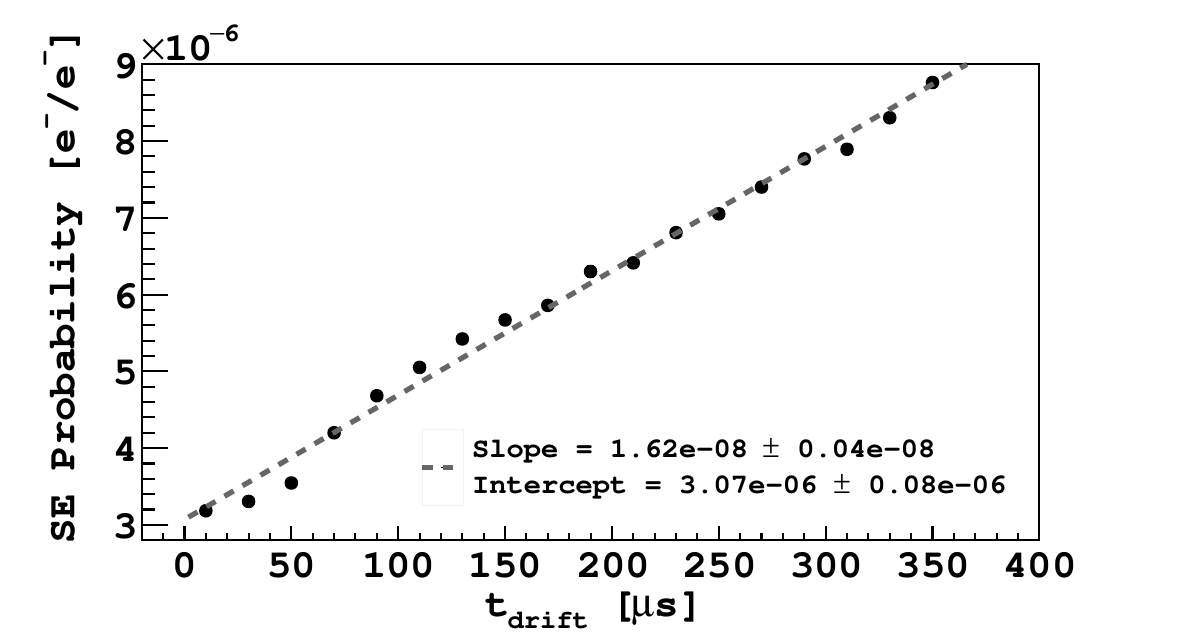}
 \caption{Total charge of all SEs following parent events, normalized by the $\STwo_\text{parent}$, as a function of parent \tdrift.}
 \label{fig:SEvTdrift}
\end{center}
\end{figure}

Figure~\ref{fig:SEvTdrift} shows the SE charge probability as a function of parent \tdrift. 
Normalizing for the parent size, the SE charge probability is estimated as $\sum_i\STwo_{\text{SE},i}/\STwo_\text{parent}$.
Note that it is calculated per unit parent charge rather than per parent event. 
The relationship between \tdrift\ and SE charge probability is well-described by a line, with slope governed by correlated SEs and $y$-intercept by uncorrelate SEs.

The linear fit gives an SE probability per unit \tdrift\ of \SI[per-mode=reciprocal]{1.62\pm0.04e-8}{e^{-}/e^{-}\per\micro\second}. 
With a \DSfLArBelowMeshElectronSpeed\ drift speed, this value corresponds to an SE probability per unit drift length of \SI[per-mode=reciprocal]{1.74(4)e-8}{e^{-}/e^{-}\per\milli\meter}.
Notably, the ratio of SE electrons to parent \STwo\ electrons is far smaller than the number of electrons lost by the finite free electron lifetime, discussed in Section~\ref{sec:purity}.

\subsection{Horizontal position correlation with previous events}

\begin{figure}[htb]
\begin{center}
\includegraphics[width=\columnwidth]{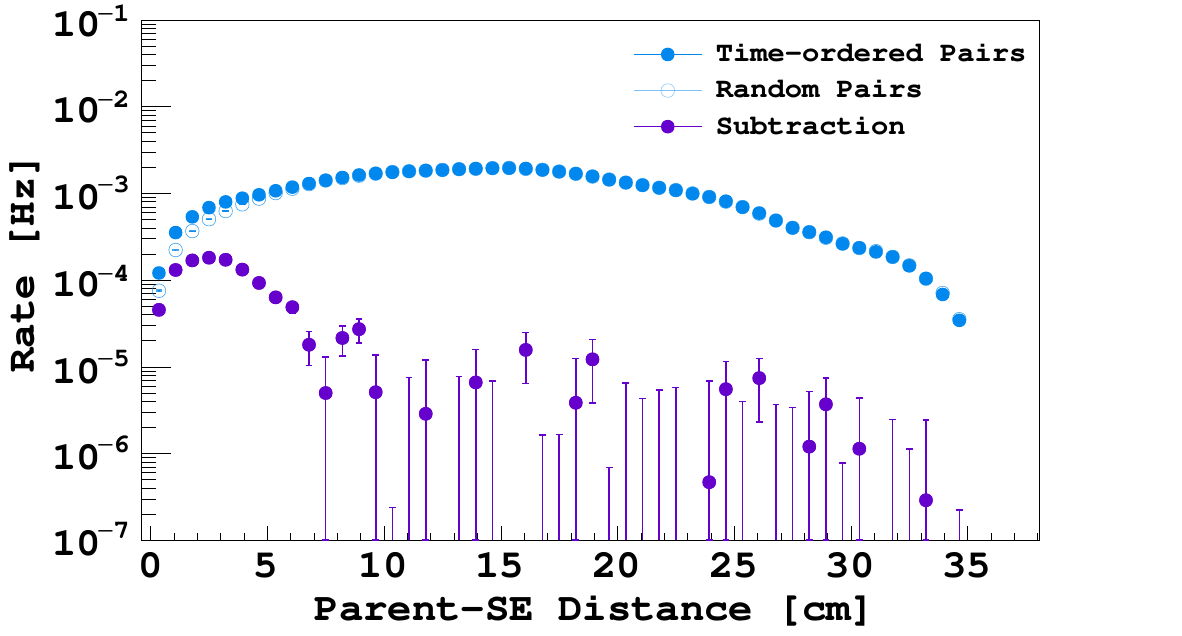}
 \caption{Rate of parent-SE pairs with a given horizontal distance, for (closed circles) correlated and (open circles) random pairs. Purple points result from subtracting random from correlated pairs.} 
 \label{fig:time_space_corr}
\end{center}
\end{figure}

To investigate if temporally-correlated SE-parent pairs are spatially correlated, the distance between SEs and parents in the horizontal plane is studied using the same transverse position reconstruction algorithm as in Ref.~\cite{Agnes:2018ep}.

Figure~\ref{fig:time_space_corr} shows the reconstructed horizontal distance between all time-ordered SEs-parents pairs within a \SI{10}{s} window, compared to that for random SE-parent pairs.
After subtracting the distribution of the random pairs from the time-ordered pairs, the temporally-correlated pairs predominantly reconstruct within \SI{5}{\cm} of each other.
These trends suggest that temporally-correlated pairs are also correlated in their position in the horizontal plane.

\subsection{Correlation with monitored hardware parameters}
\begin{figure}[htb]
\begin{center}
\includegraphics[width=\columnwidth]{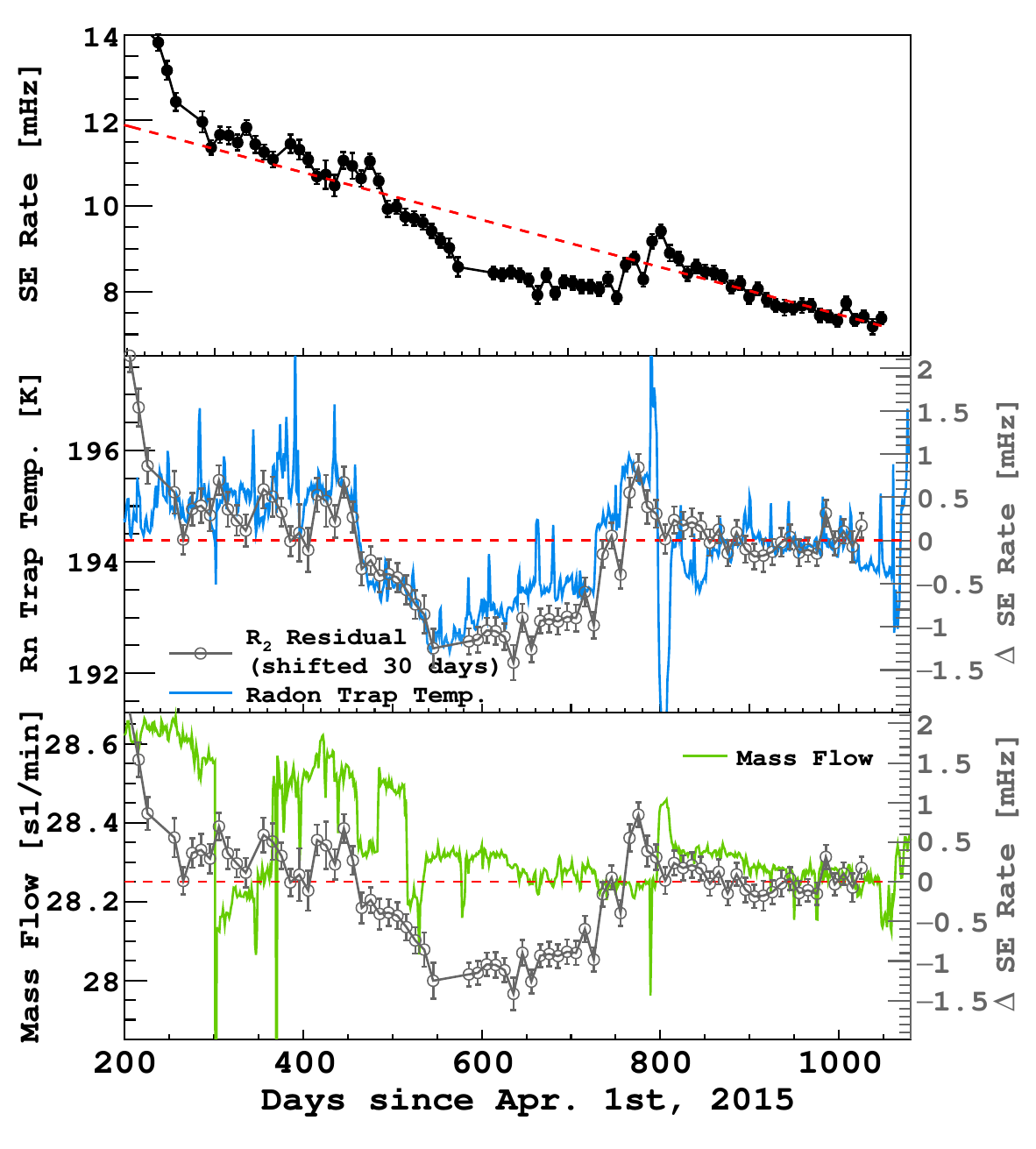}
 \caption{Top: (black) Evolution of the SE rate from the slow component $R_2$, fit with a (red) linear function. Midddle: (blue) radon trap temperature compared with (gray) the difference between the SE rate and the linear fit, shifted by \SI{30}{days}. Bottom: (green) the argon mass flow rate of the circulation, compared with (gray) the same residual rate.
 Dashed red lines are extrapolated from the linear fits over the stable period between \SIrange[range-units=single,range-phrase=--]{860}{1060}{days}.
 }
 \label{fig:Slow_Corr}
\end{center}
\end{figure}

Correlations between the SE rate and monitored detector parameters were studied to investigate the relationship between SEs and the cryogenic and purification systems' operating conditions. 

Figure~\ref{fig:Slow_Corr} shows the evolution of $R_2$ over time, along with deviations from a linear fit compared to the radon trap temperature and the circulation line mass flow, including the hot getter and radon trap.
While no correlations were seen for $R_1$, deviations of $R_2$ and the radon trap temperature correlate with each other, offset by approximately \SI{30}{days}.
No clear correlation is seen for the mass flow.
Since downward fluctuations in radon trap temperature correlate with lower $R_2$ and, for most molecules, charcoal radon traps capture more molecules at lower temperatures, these observations may indicate that the trap plays a role in the SE event rate. 
No other parameters showed an obvious correlation with the SE rate change and the mass flow rate through the hot getter and radon trap. 
The \SI{30}{day} offset is possibly related to a breakthrough time (the time it takes for a substance to traverse the radon trap)  of an impurity that is responsible for the $R_2$ component in the radon trap.

\section{Electron multiplicity in SE events}
\label{sec:ele_multi}
Characterizing SEs' \Ne distribution is important for predicting SE backgrounds in future experiments~\cite{agnesSensitivityProjectionsDualphase2022,noauthor_darkside-20k_2024}.
\begin{figure}
    \centering
    \includegraphics[width=\linewidth]{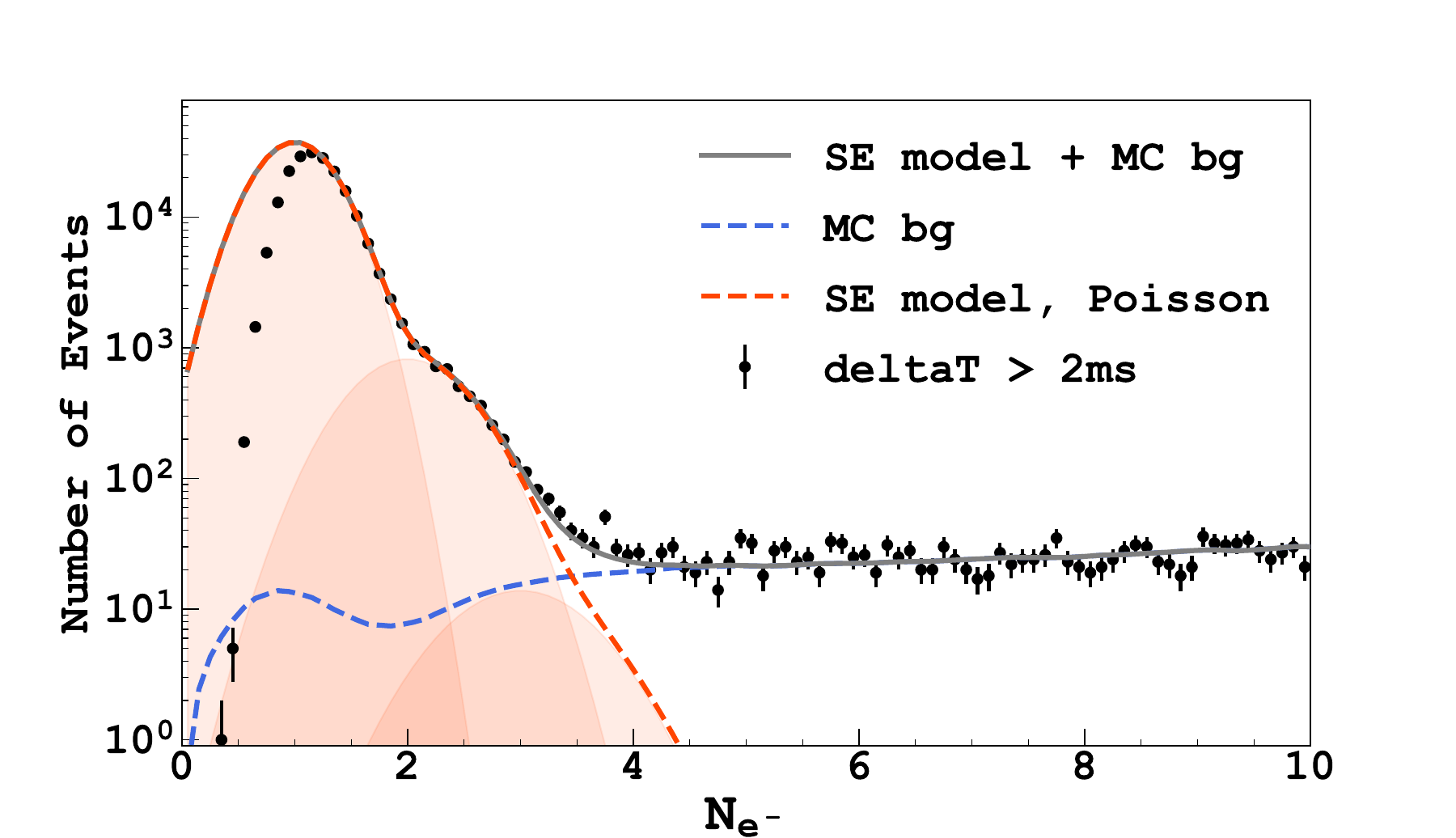}
    \caption{\Ne\ distribution fit with Monte Carlo simulations of electronic recoil backgrounds (``MC bg'', dashed blue) and SEs (dashed orange), modeled as a Poisson distribution convolved with a Gaussian (pink shaded regions for $n=1,2,3$). }
    \label{fig:se_nedist}
\end{figure}

Figure~\ref{fig:se_nedist} shows the reconstructed \Ne\ distribution observed at low energies, compared to Monte Carlo simulations of the radiological background model described in Ref.~\cite{Agnes:2018fg}. 
Above \SI{4}{\el}, the distribution is well-described by the electronic recoil background model, while it is dominated by SEs below.
The SE distribution is fit by a Poisson distribution describing the amplitude of each $n$-electrons peak, which is modeled as a Gaussian distribution with variance scaling with the number of true extracted electrons $n$:
\begin{gather}
\begin{aligned}
    \text{SE}(\Ne) &= A_\text{SE}\sum_{n=1}^{4} \text{Poisson}(n;\mu_\text{SE})\times \text{Gaus}(\Ne;n,\sigma), \\
    \sigma &= \sqrt{n}\sigma_{\text{S2}},
\end{aligned}
\end{gather}
where $\sigma$ is the electron-counting resolution, parameterized by $\sqrt{n}$ and the resolution of the \SI{1}{\el} yield, $\sigma_{\text{S2}}$.
$A_\text{SE}$ is an overall normalization factor.
$\mu_\text{SE}$ is the expected number of electrons that reconstruct in the same pulse as an SE, and the characteristic parameter of the Poisson distribution.
The model is fit to the data starting at \Ne$=1.2$, above which the electron tagging efficiency approaches \SI{100}{\percent}.
This fit converges with reduced $\chi^2= 152/134$, with 
$\mu_\text{SE}=\num{0.062\pm0.001}$ and $\sigma_{\text{S2}}=\num{0.335\pm0.001}$. 
The expected number of electrons in an SE event is given by calculating the expected number of electrons from the Poisson probability without the $n=0$ case, $\mu_\text{SE}/(1-e^{-\mu_\text{SE}})=1.031$.
The same fits are performed for different slices of $T_{diff}$, the time difference between parent and SE events. The fits confirm within statistical uncertainties that \Ne distributions are consistent over different $T_{diff}$ bins, including during the getter-off period.

\begin{figure}
    \centering
    \includegraphics[width=\linewidth]{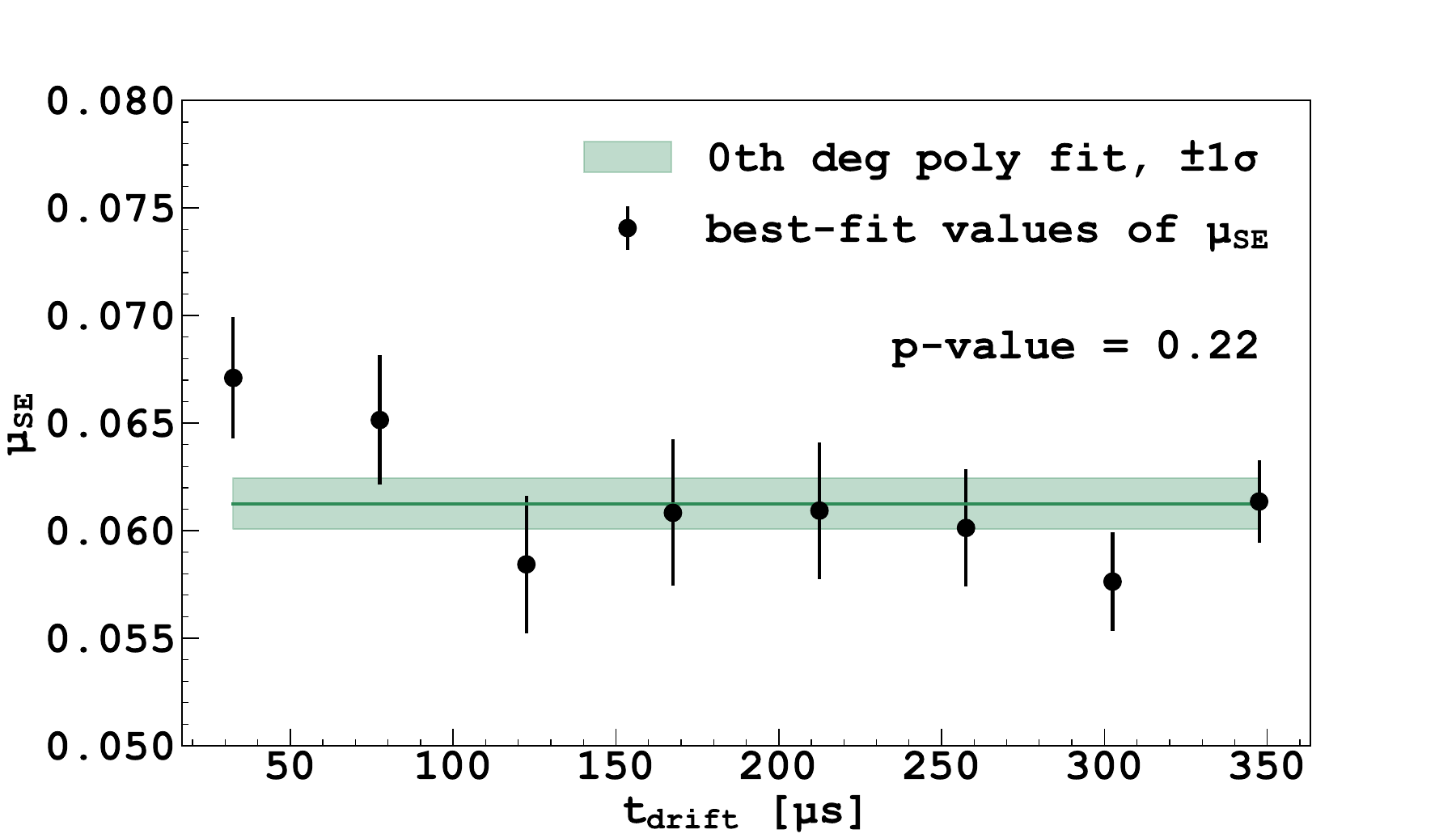}
    \includegraphics[width=\linewidth]{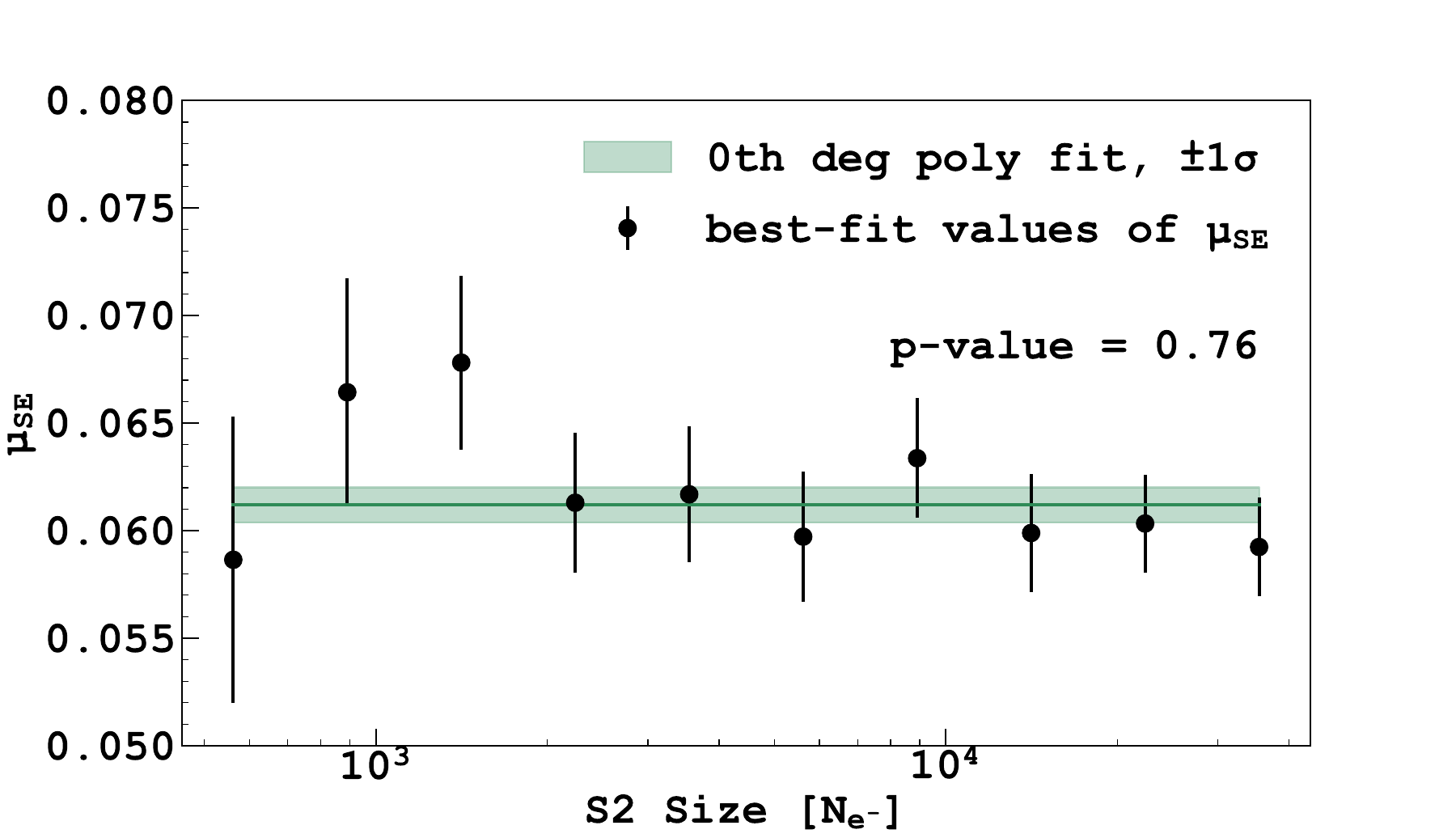}
    \caption{Stability of $\mu_{SE}$. The best values of $\mu_{SE}$ vs. parents' $t_{drift}$ (top) and vs. parent S2 size (bottom) with constant-value fits and their $\pm 1\sigma$ bands in green.}
    \label{fig:mu_se}
\end{figure}

Therefore, $\mu_\text{SE}$ does not depend on the impurity species causing SE events. The variation of $\mu_\text{SE}$ as a function of parent drift time and S2 size was also checked and is shown in Fig.~\ref{fig:mu_se}. No significant variation in $\mu_\text{SE}$ with respect to parent \tdrift or S2 size is observed; the best fit values of $\mu_\text{SE}$ have \SI{\sim5}{\%} standard deviation with average error bars of \SI{\sim6}{\%}, and exhibit no obvious trends.


A potential explanation for multiple electron emission is that \SI{128}{\nm} electroluminescence photons may produce secondary electrons by photo-ionizing the stainless steel extraction grid. The photoelectric yield of stainless steel is measured to be \SI{0.8}{\percent} at \SI{130}{\nm} and \SI{1.2}{\percent} at \SI{124}{\nm} in Ref.~\cite{laulainenPhotoelectronEmissionMetal2015}.
The grid's transparency at normal incidence is \SI{95}{\percent}, and the electroluminescence yield ranges from \SI{120}{\ph\per\el} beneath the central \PMT\ to \SI{80}{\ph\per\el} beneath the inner ring of \PMTs~\cite{zhuStudyArgonElectroluminescence2018}. Assuming that half of these UV photons are directed towards the grid,
these values predict a \SIrange[range-phrase={--}, range-units=single]{2}{4}{\percent} probability of a \SI{1}{\el} event ionizing the grid. This is consistent with the best-fit value of $\mu_\text{SE}$.


Secondary electrons reach the gas pocket \SI{\sim 1.4}{\micro\second} after the photon from the first electron strikes the extraction grid, since the \LAr\ layer extends \DSfLArAboveMeshHeight\ above it, and electrons travel with \SIrange[range-units = single,range-phrase=--]{3.1}{3.4}{\mm\per\us} 
drift speeds in the \SIrange[range-units = single,range-phrase=--]{2.8}{3.7}{\kilo\volt\per\cm} 
extraction field~\cite{LI2016160}. 
This time difference is small compared to the \SI{3.4}{\micro\second} slow component lifetime of electroluminescence~\cite{Agnes:2018hvf}, so the pulse-finding algorithm fails to identify them as a separate \STwo\ pulse.

\begin{figure}[htb]
\begin{center}
\includegraphics[width=\columnwidth]{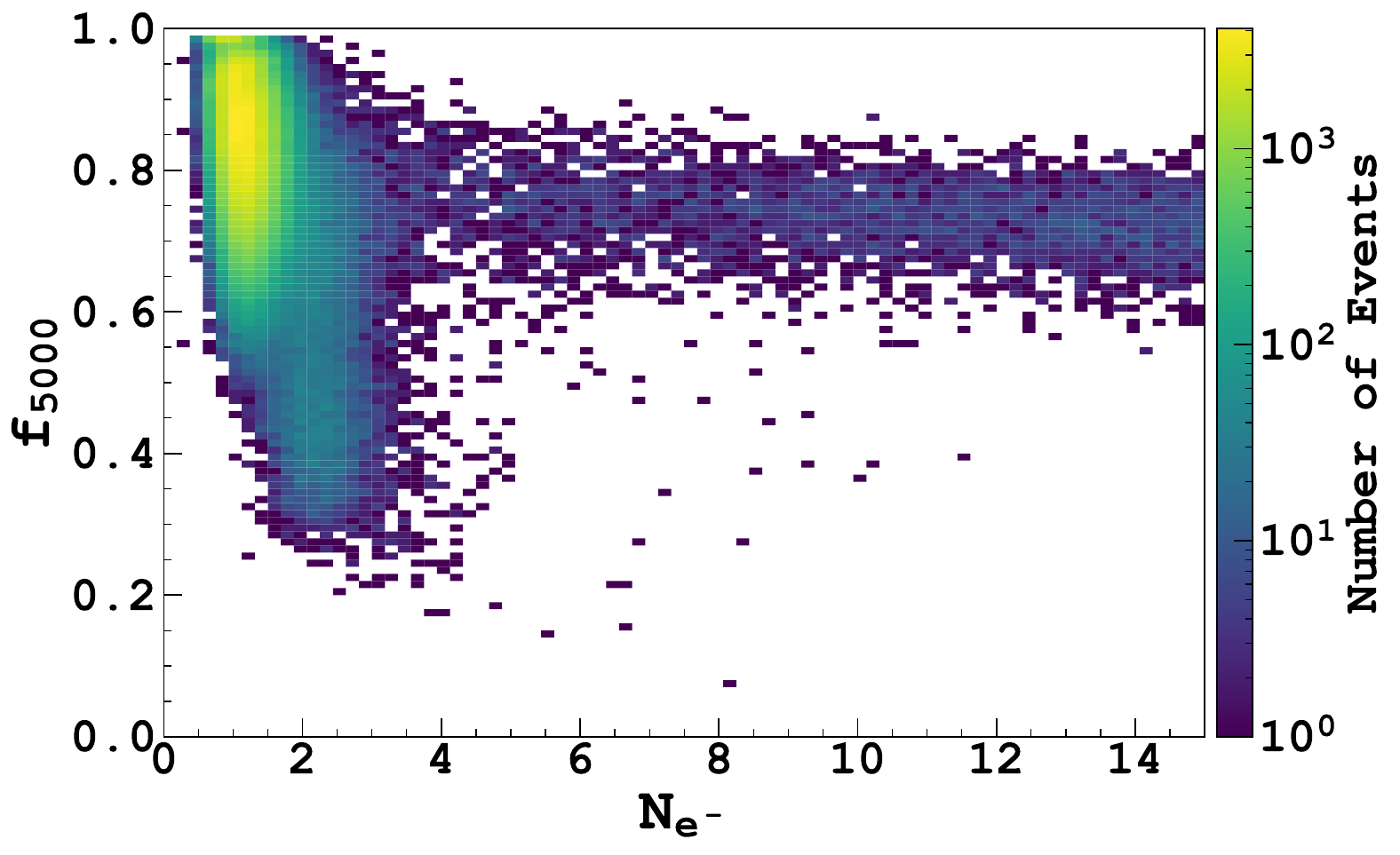}
\caption{Pulse shape parameter, $f_{5000}$, vs. \Ne. The lower $f_{5000}$ of the distribution around \Ne=2-3 indicates more spread pulses than $\Ne>4$. The z-axis is the number of events in each bin.}
 \label{fig:f5000}
\end{center}
\end{figure}

Evidence of secondary electrons delayed by $\mathcal{O}(\SI{1}{\micro\second})$ can be also be seen in SE pulse shapes. In Fig.~\ref{fig:f5000}, a pulse shape parameter, $f_{5000}$, the fraction of light in the first \SI{5}{\us} of the entire pulse, is shown as a function of \Ne. 
The population between $\Ne =$ 2 and $\Ne =$ 3 extends to lower $f_{5000}$ than the population with $\Ne>4$, showing clear distinction in the pulse shape. 
The lower mean $f_{5000}$ value for two to three electrons indicates that electrons reach the gas pocket at different times while the pulse finder reconstructs them as a single pulse. 
This behavior is seen qualitatively in recorded waveforms: two separate pulses separated by $\mathcal{O}(\SI{1}{\us})$ are observed for SEs with $\Ne>1$.
Further investigations of the pulse finder algorithm's behavior at low PE are needed to fully test this hypothesis, and is left to future work. 

\section{Discussion}
\label{sec:concl}

While there may be multiple SE sources, these observations suggest two phenomenological classes: \emph{temporally-correlated} SEs and \emph{uncorrelated} SEs. 

\subsection{Temporally-correlated SEs}
The temporally-correlated SEs appear to have a parent event with which they demonstrate strong spatial and temporal correlation (separated by an exponential time constant on a \SIrange[range-units=single,range-phrase=--]{1}{1000}{\milli\second} time scale), and the probability of observing a correlated SE scales linearly with the parent's \STwo\ size and \tdrift.
There are at least three components: a fast component of \SI{\sim5}{ms} ($\tau_1$), a slower component of \SIrange[range-phrase=--,range-units=single]{\sim40}{80}{\ms} ($\tau_2$), and a component of \SI{\sim16}{\ms} that appears when the getter is bypassed from the circulation loop. Figure~\ref{fig:Slow_Corr} shows that $R_2$ fluctuations correlate with radon trap temperature.
Those observations suggest that the cause of the correlated SEs is impurities in the \LAr\ volume, and there are at least three species causing $\tau_1$, $\tau_2$, and the component with the getter off.  
There may also be an SE population correlated with parent events by a long enough delay that parent-SE pairs are not correctly identified.




The following are observations that hint at species of the impurities. 
1)~The stable SE component with $\tau_1$ may exist primarily in the liquid and be unaffected by the gas-phase recirculation and purification system. It may not be removed by the getter or at an equilibrium concentration, given some constant source.
2)~If the population described by $R_2$ in Fig.~\ref{fig:tcorr_param} results from two impurities, the change of $\tau_2$ after \SI{200}{days} may be explained by the presence of a volatile impurity with $\tau_2\SI{\sim80}{\milli\second}$ that is removed, leaving a less volatile one with $\tau_2\SI{\sim40}{\milli\second}$.
3)~The correlation with radon trap temperature may indicate that at least one impurity enters the gas phase and is partially removed by the cold trap, since the adsorption of charcoal in the radon trap increases when it is colder.
4)~The new component that appeared when the getter is bypassed from the circulation loop may be explained by the introduction of an impurity that is volatile at \SI{87}{\kelvin} and mainly stays in the gas phase since this component was removed within a short time (\SI{36}{hour}) by the recirculation system.
5)~Based on the fact that in the getter-off runs, there is no noticeable degradation in the electron lifetime, at least the impurity causing SE events with the time constant of \SI{\sim16}{ms} in the getter-off data must be different from the impurity causing the electron lifetime loss.

Although the high electronegativity of \ce{O2} makes it a candidate explanation for $\tau_e$, it may also prevent \ce{O2} from releasing electrons as SEs.
%
%
\ce{N2} may be a good candidate consistent with some of the above observations, as it boils at \SI{77}{\kelvin} and captures electrons in LAr with an attachment coefficient \SI{\sim1}{\percent} that of \ce{O2}~\cite{Swan:1965ha}.
Also, \ce{N2} has slightly higher electron affinity of \SI{-0.07}{eV} than \ce{Ar} (\SI{-0.1}{eV}) and in \ce{Ar} media, the effective electron affinity of \ce{N2} could be positive.
However, the triplet lifetime study suggests that the \ce{N2} contamination was \SI{<1}{\ppm} level as shown in Sec.~\ref{sec:pulseshape}.
The long lifetime of metastable anions that produce SEs, according to this mechanism, can potentially be explained by long resonant autodetachment times typically seen in diatomic molecular anions, from long-lived vibrational states (prolonged at low temperatures)~\cite{millikanSystematicsVibrationalRelaxation2004} or small energy level gaps~\cite{acharyaVibrationinducedElectronDetachment1984a}.
Similar mechanisms make \ce{OH-} or residual moisture plausible SE sources from non-volatile impurities.
TPB is another candidate: it is soluble in LAr~\cite{asaadiEmanationBulkFluorescence2019} and scintillates with $\mathcal{O}(\SI{1}{\milli\second})$ lifetimes~\cite{Stanford:2018un} (implying the existence of long-lived molecular excited states consistent with lower rates of inter-system crossing in TPB's molecular structure), and similar benzenoid molecules have long-lived anionic states~\cite{simonsMolecularAnions2008,fieldVeryLowenergyElectron2001}.

The hypothesis that impurities cause the temporally-correlated SEs is supported by the observation. However, to confirm the hypothesis, identify impurities, and develop means to reduce SE events in future experiments, the SE rate needs to be studied with spiked impurity concentrations in a dedicated system.



\subsection{Uncorrelated SEs} 

For uncorrelated SEs, no parent has been identified within the \SI{\sim1}{\second} timescales accessible in this study.
%
This component is either from spontaneous causes or correlated SE components on a time scale \SI{>1}{s}. 
%
Figure~\ref{fig:dRdT_long} shows that a component extends out to \SI{\approx10}{\second} timescales at a rate comparable to $R_1$ and $R_2$.
Due to its long lifetime, sufficient statistics are only achieved with a long period, making it hard to study this component in the same way as the other shorter components.
It is possible that all or most of the so-called uncorrelated SEs actually do have a parent event that has not been identified.
This explanation is supported by the observation that the uncorrelated SE rate in Fig.~\ref{fig:rate_v_Ne} is proportional to the rate of ionization. 
One of the possible mechanisms for this long-time correlation is that negative ions drift to the extraction region and release electrons. 
Some parents may also be missed by the selection criteria, though this is unlikely to be a dominant source of uncorrelated SEs given the high acceptance of these criteria and the low SE production rate for the low-energy events that fail cuts, as shown in Fig.~\ref{fig:SEvS2}.

The upper bound on the constant component of the best-fit line to the ``uncorrelated'' SE rate in Fig.~\ref{fig:rate_v_Ne} implies that any SE production mechanisms truly unrelated to ionization of the \LAr\ must have a rate \SI{<7.5}{\milli\hertz}.
Such mechanisms can include spontaneous electron emission from the grid~\cite{tomasStudyMitigationSpurious2018,baileyDarkMatterSearches2016}.
While spontaneous emission of electrons trapped at the gas/liquid interface has been proposed as an SE-production mechanism in \LXe\ experiments~\cite{akeribInvestigationBackgroundElectron2020,Sorensen:2017kpl,Sorensen:2017ymt}, this mechanism is unlikely to explain SEs in \DSf, as the electron-transparency of the liquid surface is much higher in \LAr\ than \LXe~\cite{gushchinElectronEmissionCondensed1979}, and the corresponding fast component of trapped-electron emission is not observed~\cite{cohenTheoryHotElectrons1967}.

\subsection{Comparisons with Xenon Experiments}

Dual-phase liquid xenon experiments have also observed SE signals. A study of SEs in the XENON1T experiment concluded that they were both temporally and spatially correlated with preceding high-energy events~\cite{Aprile_2021}. 
The time-correlated SEs in XENON1T follow a power law. On the other hand, in \DSf\, at least one component of SEs released following a parent event shows a clear exponential dependence, described by $\tau_2$.
This suggests that at least one different mechanism for SEs exists in \LAr\ compared to \LXe. 
Power-law exponents differed between \SI{1}{\el} SEs and multi-electron SEs, indicating that those with $\Ne > 1$ were not simply the result of pile-up, but were rather produced by some mechanism capable of trapping and releasing more than one delayed electron. 

A study of SEs in the LUX detector~\cite{akeribInvestigationBackgroundElectron2020} found that SEs were primarily caused by two populations: bursts of electrons trapped under the liquid surface, and drifting electrons trapped by impurities. 
While evidence for the former mechanism is not observed in \DSf, delayed release from impurities is identified as a potential source of correlated SEs in this work. Similarly to the XENON1T experiment, the delay times between SEs and their parents in LUX followed a power-law. 

Extended electron tails on time scales shorter than the drift length have not been observed immediately following \STwo\ pulses in \DSf; the \STwo\ pulse shape is consistent with the expected electroluminescence pulse shape~\cite{Agnes:2018hvf} and shows no evidence of electrons delayed at the liquid/gas interface, contrary to what has been reported in xenon \TPCs~\cite{Sorensen:2017kpl,Sorensen:2017ymt,akeribInvestigationBackgroundElectron2020}.

\section{Summary}
\label{sec:summary}
Events with fewer than four extracted electrons behave differently from other S2-only events. Some fraction of these SE events (\SIrange[range-phrase=--,range-units=single]{\sim30}{70}{\percent} of SEs, Fig.~\ref{fig:corr_uncorr_rate}) are likely electrons released from impurities. 
These events are correlated to preceding events in time, position, and energy. The rate of this correlated component increases with the parents' \STwo\ size and drift time. From these correlations, a trapping probability of \SI[per-mode=reciprocal]{1.74e-8}{\per\mm} drift length is estimated with the getter on. 
A study of temporal correlations suggests that the timescale at which these trapped electrons are released from impurities is dominated by an exponential distribution with at least two components: a fast component of \SI{\sim5}{ms} ($\tau_1$), a slower component of \SIrange[range-phrase=--,range-units=single]{\sim40}{80}{\ms} ($\tau_2$), and a component of \SI{\sim16}{\ms} that appears when the getter is turned off. 
The change in the value of $\tau_2$ over time may indicate that this time constant describes two different components with changing relative abundances. We also note the existence of a longer-lived (\SI{>1}{s}) component not described by this exponential model. 
Observations are consistent with the hypothesis that uncorrelated SE events are correlated SE components with a lifetime of \SI{>1}{s}. 

The electron multiplicity in SE events is well described by a Poisson distribution characterized by a mean number of electrons $\mu_\text{SE}$. The $\mu_\text{SE}$ does not depend on the impurity species, parent's drift time, and parent's S2 size. The $f_{5000}$ parameter in Fig.~\ref{fig:f5000} indicates that the electrons in the SEs events with 2- and 3-electron multiplicity are not extracted into the gas pocket simultaneously. Those observations are consistent with the hypothesis that electroluminescence photons ionize the extraction grid, resulting in secondary electrons. If the photo-ionization yield of an extraction grid could be suppressed, the analysis threshold, which is limited by the leakage of SE events currently, might be lowered in future experiments.
Confirmation of this mechanism will allow SEs with $\Ne>1$ to be properly included in background models, allowing searches for low-mass DM at even lower thresholds than current generation experiments, and thus to probe lower masses and cross sections.

\section{Acknowledgment}
\small{
The DarkSide Collaboration offers its deep gratitude to LNGS and its staff for their invaluable technical and logistical support. 
The authors also thank the Fermilab Particle Physics, Scientific, and Core Computing Divisions.
Construction and operation of the DarkSide-50 detector was supported by the U.S. National Science Foundation (NSF) (Grants No. PHY-0919363, No. PHY-1004072, No. PHY-1004054, No. PHY-1242585, No. PHY-1314483, No. PHY-1314501, No. PHY-1314507, No. PHY-1352795, No. PHY-1622415, and associated collaborative Grants No. PHY-1211308 and No. PHY-1455351), 
the Italian Istituto Nazionale di Fisica Nucleare, 
the U.S. Department of Energy (Contracts No. DE-FG02-91ER40671, No. DEAC02-07CH11359, No. DE-AC05-76RL01830, and No. DE-SC0025540), the Polish NCN (Grant No. UMO-2023/51/B/ST2/02099) 
 and the Polish Ministry for Education and Science (Grant No. 6811/IA/SP/2018). 
 We also acknowledge financial support from the French Institut National de Physique Nucl\'eaire et de Physique des Particules (IN2P3), the IN2P3-COPIN consortium (Grant No. 20-152), and the UnivEarthS LabEx program (Grants No. ANR-10-LABX-0023 and No. ANR-18-IDEX-0001), 
 from the S{\~a}o Paulo Research Foundation (FAPESP) (Grant No. 2017/26238-4 and 2021/11489-7), 
 from the Interdisciplinary Scientific and Educational School of Moscow University “Fundamental and Applied Space Research,” 
 from the Program of the Ministry of Education and Science of the Russian Federation for higher education establishments, Project No. FZWG-2020-0032 (2019-1569), 
 the FNP IRA programmes: AstroCeNT (MAB/2018/7), funded from the ERDF, and Astrocent (FENG.02.01-IP.05-A015/25) co-financed by the European Union under FENG 2021–2027; and Teaming for Excellence grant Astrocent Plus (101137080) funded by the European Union with complementary national funding from the MNiSW (MNiSW/2025/DIR/811),
 the European Union’s Horizon 2020 research and innovation program under Grant Agreement No. 952480 (DarkWave), the National Science Centre, Poland (2021/42/E/ST2/00331), 
 and from the Science and Technology Facilities Council, United Kingdom.
  I. Albuquerque was partially supported by the Brazilian Research Council (CNPq).
}

\bibliographystyle{ds}
\bibliography{ds,wal}

\onecolumn
\textbf{The DarkSide-50 Collaboration}

P.~Agnes\thanksref{l_AQGSSI}\textsuperscript{,}\thanksref{l_AQLNGS}\nolinebreak,
I.~F.~Albuquerque\thanksref{l_USP}\nolinebreak,
T.~Alexander\thanksref{l_PNNL}\nolinebreak,
A.~K.~Alton\thanksref{l_Augustana}\nolinebreak,
M.~Ave\thanksref{l_USP}\nolinebreak,
H.~O.~Back\thanksref{l_PNNL}\nolinebreak,
G.~Batignani\thanksref{l_PIUniPHY}\textsuperscript{,}\thanksref{l_PIINFN}\nolinebreak,
E.~Berzin\thanksref{l_Princeton}\textsuperscript{,}\thanksref{l_Stanford}\nolinebreak,
K.~Biery\thanksref{l_FNAL}\nolinebreak,
V.~Bocci\thanksref{l_RMUnoINFN}\nolinebreak
% ...

\begin{enumerate}
\item{\label{l_AQGSSI}\AQGSSI}
\item{\label{l_AQLNGS}\AQLNGS}
\item{\label{l_USP}\USP}
\item{\label{l_PNNL}\PNNL}
\item{\label{l_Augustana}\Augustana}
\item{\label{l_PIUniPHY}\PIUniPHY}
\item{\label{l_PIINFN}\PIINFN}
\item{\label{l_FNAL}\FNAL}
\item{\label{l_RMUnoINFN}\RMUnoINFN}
\item{\label{l_CAINFN}\CAINFN}
\item{\label{l_GEUni}\GEUni}
\item{\label{l_GEINFN}\GEINFN}
\item{\label{l_RMTreUni}\RMTreUni}
\item{\label{l_RMTreINFN}\RMTreINFN}
\item{\label{l_CAUniPHY}\CAUniPHY}
\item{\label{l_Princeton}\Princeton}
\item{\label{l_london}\london}
\item{\label{l_VTech}\VTech}
\item{\label{l_MSU}\MSU}
\item{\label{l_MIUni}\MIUni}
\item{\label{l_MIINFN}\MIINFN}
\item{\label{l_RMUnoUni}\RMUnoUni}
\item{\label{l_Petersburg}\Petersburg}
\item{\label{l_UMass}\UMass}
\item{\label{l_NAUniPHY}\NAUniPHY}
\item{\label{l_NAINFN}\NAINFN}
\item{\label{l_APC}\APC}
\item{\label{l_LPNHE}\LPNHE}
\item{\label{l_LNFINFN}\LNFINFN}
\item{\label{l_Kurchatov}\Kurchatov}
\item{\label{l_IHEP}\IHEP}
\item{\label{l_ENUniCEE}\ENUniCEE}
\item{\label{l_CTLNS}\CTLNS}
\item{\label{l_CPPM}\CPPM}
\item{\label{l_Houston}\Houston}
\item{\label{l_SDakota}\SDakota}
\item{\label{l_AstroCeNT}\AstroCeNT}
\item{\label{l_JINR}\JINR}
\item{\label{l_Belgorod}\Belgorod}
\item{\label{l_UCRiverside}\UCRiverside}
\item{\label{l_MEPhI}\MEPhI}
\item{\label{l_UCLA}\UCLA}
\item{\label{l_Hawaii}\Hawaii}
\item{\label{l_Oxford}\Oxford}
\item{\label{l_PGUniCBB}\PGUniCBB}
\item{\label{l_PGINFN}\PGINFN}
\item{\label{l_UCDavis}\UCDavis}
\item{\label{l_Krakow}\Krakow}
\item{\label{l_Manchester}\Manchester}
\item{\label{l_Washington}\Washington}
\item{\label{l_UCAS}\UCAS}
\item{\label{l_Stanford}\Stanford}
\item{\label{l_Williams}\Williams}
\end{enumerate}

\end{document}